\pdfoutput=1

\documentclass[useAMS,usenatbib,fleqn]{mnras}

\usepackage{graphicx,amssymb}
\usepackage{epsf,psfig}
\usepackage{natbib}
\usepackage{amsmath}

\title[Orbital and escape dynamics in barred galaxies - I. The 2D system]{Orbital and escape dynamics in barred galaxies - I. The 2D system}

\author[Ch. Jung \& E. E. Zotos]{Christof Jung$^1$\thanks{E-mail:jung@fis.unam.mx} and Euaggelos E. Zotos$^2$\thanks{E-mail: evzotos@physics.auth.gr} \\
$^1$ Instituto de Ciencias F\'{i}sicas, Universidad Nacional Aut\'{o}noma de M\'{e}xico
Av. Universidad s/n, 62251 Cuernavaca, Mexico \\
$^2$ Department of Physics, School of Science, Aristotle University of Thessaloniki,
GR-541 24, Thessaloniki, Greece
}

\begin{document}

\date{Accepted 2016 January 18. Received 2016 January 18; in original form 2015 November 19}

\pubyear{2016} \volume{457} \pagerange{2583--2603}

\setcounter{page}{2583}

\maketitle

\label{firstpage}

\begin{abstract}
In this paper we use the two-dimensional (2D) version of a new analytical gravitational model in order to explore the orbital as well as the escape dynamics of the stars in a barred galaxy composed of a spherically symmetric central nucleus, a bar, a flat disk and a dark matter halo component. A thorough numerical investigation is conducted for distinguishing between bounded and escaping motion. Furthermore bounded orbits are further classified into non-escaping regular and trapped chaotic using the Smaller ALingment Index (SALI) method. Our aim is to determine the basins of escape through the two symmetrical escape channels around the Lagrange points $L_2$ and $L_3$ and also to relate them with the corresponding distribution of the escape rates of the orbits. We integrate initial conditions of orbits in several types of planes so as to obtain a more complete view of the overall orbital properties of the dynamical system. We also present evidence that the unstable manifolds which guide the orbits in and out the interior region are directly related with the formation of spiral and ring stellar structures observed in barred galaxies. In particular, we examine how the bar's semi-major axis determines the resulting morphologies. Our numerical simulations indicate that weak barred structures favour the formation of $R_1$ rings or $R_1'$ pseudo-rings, while strong bars on the other hand, give rise to $R_1R_2$ and open spiral morphologies. Our results are compared with earlier related work. The escape dynamics and the properties of the manifolds of the full three-dimensional (3D) galactic system will be given in an accompanying paper.
\end{abstract}


\section{Introduction}
\label{intro}

Bars are linear extended stellar structures located in the central regions of disk-shaped galaxies. Observations reveal that many spiral galaxies contain bars in their central regions. A long time ago de Vaucouleurs \citep{dV63} established that about one third of the observed disk galaxies do not contain a bar, one third have intermediate or undeterminable types of bars, while the remaining third display strong bar properties \citep[e.g.,][]{Ee00,SRSS03}. Recent observations indicate that the fraction of spiral galaxies with barred structures declines with increasing redshift \citep[e.g.,][]{MNH11,MM13,SEE08}. The dynamical reason for the occurrence of galactic bars is generally thought to be the result of a density wave radiating from the center of the galaxy whose effects reshape the orbits of the stars located in the interior region. As time goes on this effect causes stars to orbit further out and thereby creating a self stabilizing bar structure \citep[e.g.,][]{BC02}.

In a recent paper \citet{JZ15} (hereafter Paper I) we introduced a new analytical gravitational model describing the properties of stars in a barred galaxy with a central spherically symmetric nucleus with an additional flat disk. The main advantage of the new bar potential is its relative simplicity. Until Paper I the most realistic bar potential is the Ferrers' triaxial model \citep{F77} \citep[e.g.,][]{P84}. However, the corresponding potential is too complicated, while it is not known in a closed form. On this basis, our new simpler, yet realistic potential, has a clear advantage on the performance speed of the numerical calculations in comparison with the Ferrers' potential.

The problem of escapes is a classical problem in open Hamiltonian nonlinear systems \citep[e.g.,][]{AVS01,AS03,AVS09,BSBS12,Z14b,Z15a} as well as in dynamical astronomy \citep[e.g.,][]{BTS96,BST98,dML00,Z15b,Z15c,Z15e}. A central topic of the present paper is the escape of stars from the interior potential hole over the potential saddles. In this sense we are dealing with an open Hamiltonian system where for energies above the escape threshold the energy shell is noncompact and where at least a part of the orbits really explores an infinite part of the position space. In addition we study Hamiltonian dynamics which is time reversal invariant. Therefore, if we continue star orbits into the past, then a large majority of the orbits escapes also in past direction. In this sense, most of the orbits enter in the past, stay in the potential interior for a finite time only and disappear in the future again into the exterior, they are scattering orbits. If escaping orbits show complicated behaviour then it can only be transient chaos where they follow the chaotic dynamics for a finite time only \citep{LT11}. In such a dynamics we also have localized orbits, in particular unstable periodic orbits which are the skeleton of the chaotic invariant set. These unstable orbits have their stable and unstable invariant manifolds and if these invariant manifolds have transverse intersections, then the existence of horseshoes in the Poincar\'e map is guaranteed (for a very nice pictorial explanation of all these mathematical concepts see \citet{AS92}).

\citet{EP14} (hereafter Paper II) used a simple analytical barred galaxy model \citep{Z12} in order to investigate the escape properties of the dynamical system. They managed to locate the basins of escape and to relate them with the corresponding escape times of the orbits. They found the vast majority of both the configuration and the phase space is covered by initial conditions of orbits which do not escape within the predefined time interval of the numerical integration, while basins of escape are mainly confined near the saddle Lagrange points. In our paper we shall follow similar numerical techniques in order to unveil the escape dynamics of this new barred galaxy model.

Over the years a huge amount of work has been devoted to the issue of chaotic scattering. In this situation, a test particle coming from infinity approaches and then scatters off a potential. This phenomenon is well explored as well interpreted from the viewpoint of chaos theory \citep[e.g.,][]{BGOB88,BOG89,BGO90,JP89,JMS95,JLS99,LFO91,LGB93,LMG00}. In the galaxy true scattering dynamics is not realized since stars are formed in its interior region. However the stars which eventually escape over the potential saddles can be considered to perform half-scattering. And in this sense a good knowledge of the scattering dynamics of a system also provides the necessary information to understand the escape dynamics. Just as a mathematical exercise we can imagine the past continuation of the orbit also before the formation of the star.

Usually in open Hamiltonian systems there exists a small number of outermost elements of the chaotic invariant set whose invariant manifolds trace out the horseshoe construction. They sit over the outermost saddles of the effective potential of the system. In celestial mechanics they are frequently called Lyapunov orbits. The invariant manifolds of these outermost elements direct the flow over the saddle, i.e. they determine how general orbits enter the potential interior and leave it again. Thereby they also create the fractal structure seen in scattering functions (for the explanation of a typical example see \citet{JS87}). In addition the stable manifolds of the outermost localized orbits form the boundaries of the various basins of escape and the shape of the unstable manifolds of the Lyapunov orbits is responsible for the ring and spiral structures found in the outer parts of barred galaxies \citep[e.g.,][]{ARGM09,ARGBM09,ARGBM10,ARGM11,RGMA06,RGAM07}. All these properties are a strong justification for a detailed description of the Lyapunov orbits in our system. We shall also discuss the formation of different stellar morphologies according to the dynamical properties of the barred galaxy model.

The structure of the article is as follows: In Section \ref{galmod} we present a description of the main properties of our new barred galaxy model. In the following Section, we explore how the bar's semi-major axis influences the orbital properties of the system. Section \ref{escdyn} contains a thorough and systematic numerical investigation thus revealing the overall escape dynamics. In Section \ref{loman} we link the morphologies of the invariant manifolds with the properties of the bar, while in Section \ref{fate} we discuss the fate of escaping stars and how they can form ring and spiral structures. Our paper ends with Section \ref{disc}, where the discussion and the conclusions of this work are given.

\section{Description of the dynamical model}
\label{galmod}

In Paper I we introduced a new realistic three-dimensional (3D) dynamical model for the description of barred galaxies following the three component model presented in \citet{P84}. The total potential, $\Phi_{\rm t}(x,y,z)$, was composed of three components: a central spherical component $\Phi_{\rm n}$, a bar potential $\Phi_{\rm b}$ and a flat disk $\Phi_{\rm d}$. In this paper we add a fourth component, a spherical dark matter halo $\Phi_{\rm h}$, in order to obtain a decent asymptotic behaviour for large distances from the galactic center.

The Plummer potential \citep{BT08} is used for describing the spherically symmetric nucleus
\begin{equation}
\Phi_{\rm n}(x,y,z) = - \frac{G M_{\rm n}}{\sqrt{x^2 + y^2 + z^ 2 + c_{\rm n}^2}},
\label{Vn}
\end{equation}
where $G$ is the gravitational constant, while $M_{\rm n}$ and $c_{\rm n}$ are the mass and the scale length of the nucleus, respectively. Here we must clarify that potential (\ref{Vn}) is not intended to represent a compact object (e.g., a black hole), but a dense and massive bulge. On this basis, no relativistic effects are included.

For the description of the galactic bar we use the new potential
\begin{align}
\Phi_{\rm b}(x,y,z) &= \frac{G M_{\rm b}}{2a}\left[\sinh^{-1} \left( \frac{x-a}{d} \right) - \sinh^{-1} \left( \frac{x+a}{d} \right) \right] = \nonumber \\
&= \frac{G M_{\rm b}}{2a} \ln \left( \frac{x-a+\sqrt{(x-a)^2 + d^2}} {x+a+\sqrt{(x+a)^2 + d^2}} \right),
\label{Vb}
\end{align}
where $d^2 = y^2 + z^2 + c_b^2$, $M_{\rm b}$ is the mass of the bar, $a$ is the length of the semi-major axis of the bar, while $c_{\rm b}$ is its scale length (For more details regarding the development of the bar potential see Paper I).

A Miyamoto-Nagai potential \citep{MN75} is deployed for modelling the flat disk
\begin{equation}
\Phi_{\rm d}(x,y,z) = - \frac{G M_{\rm d}}{\sqrt{x^2 + y^2 + \left(k + \sqrt{h^2 + z^ 2}\right)^2}},
\label{Vd}
\end{equation}
where $M_{\rm d}$ is the mass of the disk, while $k$ and $h$ are the horizontal and vertical scale lengths of the disk, respectively.

Dark matter haloes have a variety of shapes, i.e., spherical, biaxial or triaxial \citep[see e.g.,][] {CZ10,CZ11,ITN00,MSW02,OM00,OHD01,PSS02,SKS02,WBP02,Z14a}. In this investigation we adopt a spherical dark matter halo using a Plummer potential
\begin{equation}
\Phi_{\rm h}(x,y,z) = - \frac{G M_{\rm h}}{\sqrt{x^2 + y^2 + z^ 2 + c_{\rm h}^2}},
\label{Vh}
\end{equation}
where $M_{\rm h}$ and $c_{\rm h}$ are the mass and the scale length of the dark matter halo, respectively.

The galactic bar follows a clockwise rotation around the $z$-axis at a constant angular velocity $\Omega_{\rm b}$. Therefore we describe the dynamics in the corresponding rotating frame where the semi-major axis of the bar points into the $x$ direction, while its intermediate axis points into the $y$ direction. The potential in the rotating frame of reference (known as the effective potential) is
\begin{equation}
\Phi_{\rm eff}(x,y,z) = \Phi_{\rm t}(x,y,z) - \frac{1}{2}\Omega_{\rm b}^2 \left(x^2 + y^2 \right).
\label{Veff}
\end{equation}

\begin{figure}
\begin{center}
\includegraphics[width=\hsize]{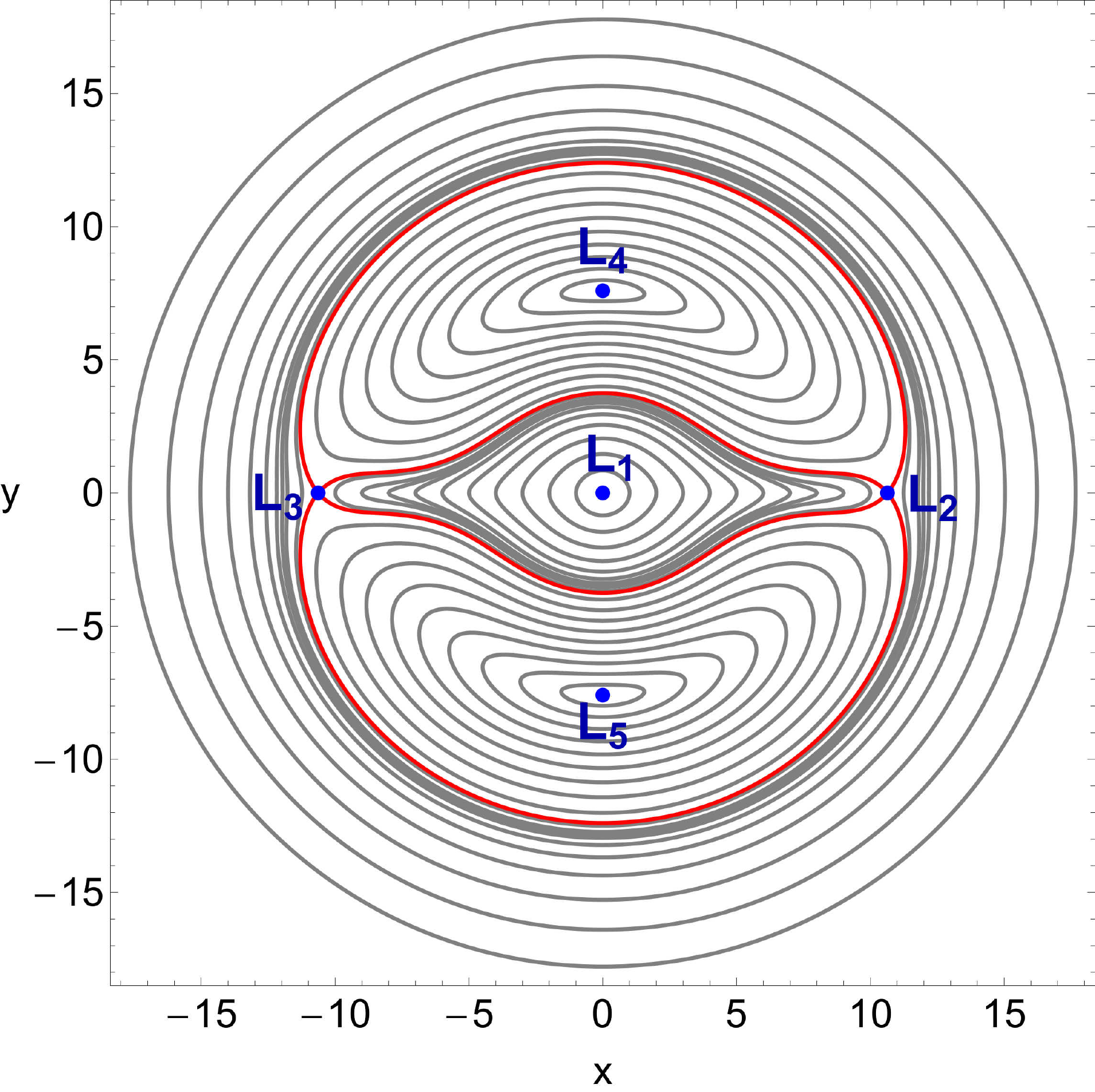}
\end{center}
\caption{The isoline contours of the effective potential in the $(x,y)$-plane for $z = 0$ for the Standard Model $(a = 10)$. Included are the five Lagrange points. The isoline contours corresponding to the critical energy of escape $E(L_2)$ are shown in red. (\textit{For the interpretation of references to color in this figure caption and the corresponding text, the reader is referred to the electronic version of the article.})}
\label{isoc}
\end{figure}

\begin{figure*}
\centering
\resizebox{\hsize}{!}{\includegraphics{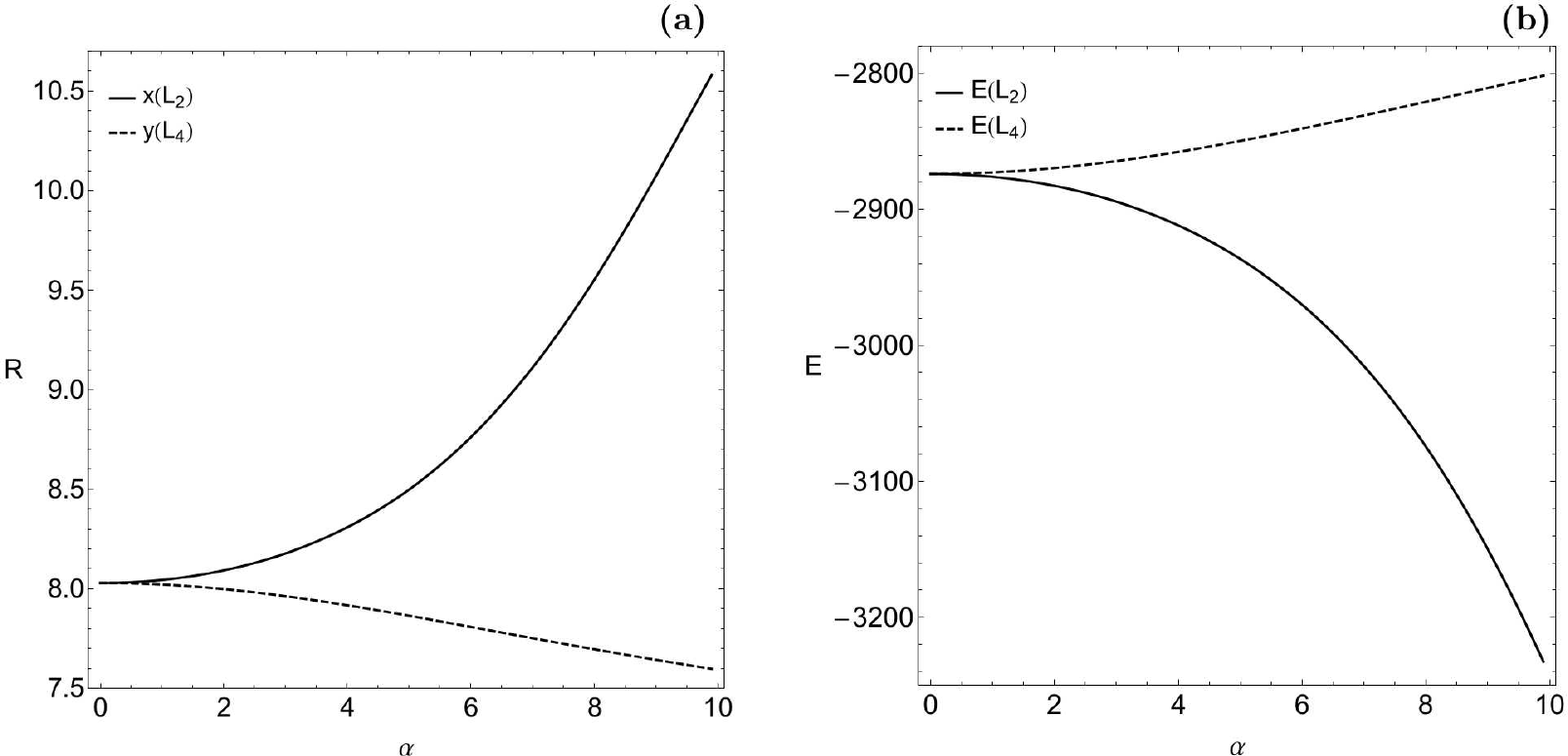}}
\caption{Evolution of (a-left): the position $(R^2 = x^2 + y^2)$ of the Lagrange points $L_2$ and $L_4$ and (b-right): the critical Jacobi values as a function of the bar's semi-major axis $a$.}
\label{theor}
\end{figure*}

We use a system of galactic units, where the unit of length is 1 kpc, the unit of mass is $2.325 \times 10^7 {\rm M}_\odot$ and the unit of time is $0.9778 \times 10^8$ yr (about 100 Myr). The velocity unit is 10 km/s, the unit of angular momentum (per unit mass) is 10 km kpc s$^{-1}$, while $G$ is equal to unity. The energy unit (per unit mass) is 100 km$^2$s$^{-2}$, while the angle unit is 1 radian. In these units, the values of the involved parameters are: $M_{\rm n} = 400$, $c_{\rm n} = 0.25$, $M_{\rm b} = 3500$, $a = 10$, $c_{\rm b} = 1$, $M_{\rm d} = 7000$, $k = 3$, $h = 0.175$, $M_{\rm h} = 20000$, $c_{\rm h} = 20$, and $\Omega_{\rm b} = 4.5$. This set of the values of the dynamical parameters defines the Standard Model (SM). In our numerical investigation only the value of the bar's semi-major axis will be varying in the interval $a \in [0,10]$, while the values of all the other dynamical parameters will remain constant according to SM.

The Hamiltonian which governs the motion of a test particle (star) with a unit mass $(m = 1)$ in our rotating barred galaxy model is \begin{equation}
H = \frac{1}{2} \left(p_x^2 + p_y^2 + p_z^2 \right) + \Phi_{\rm t}(x,y,z) - \Omega_{\rm b} L_z = E,
\label{ham}
\end{equation}
where $p_x$, $p_y$ and $p_z$ are the canonical momenta per unit mass, conjugate to $x$, $y$ and $z$ respectively, $E$ is the numerical value of the Jacobi integral, which is conserved, while $L_z = x p_y - y p_x$ is the angular momentum along the $z$ direction.

The corresponding equations of motion are
\begin{align}
\dot{x} &= p_x + \Omega_{\rm b} y, \nonumber \\
\dot{y} &= p_y - \Omega_{\rm b} x, \nonumber \\
\dot{z} &= p_z, \nonumber \\
\dot{p_x} &= - \frac{\partial \Phi_{\rm t}}{\partial x} + \Omega_{\rm b} p_y, \nonumber \\
\dot{p_y} &= - \frac{\partial \Phi_{\rm t}}{\partial y} - \Omega_{\rm b} p_x, \nonumber \\
\dot{p_z} &= - \frac{\partial \Phi_{\rm t}}{\partial z},
\label{eqmot}
\end{align}
where the dot indicates the derivative with respect to the time.

The set of the variational equations which governs the evolution of a deviation vector ${\bf{w}} = (\delta x, \delta y, \delta z, \delta p_x, \delta p_y, \delta p_z)$ is
\begin{align}
\dot{(\delta x)} &= \delta p_x + \Omega_{\rm b} \delta y, \nonumber \\
\dot{(\delta y)} &= \delta p_y - \Omega_{\rm b} \delta x, \nonumber \\
\dot{(\delta z)} &= \delta p_z, \nonumber \\
(\dot{\delta p_x}) &=
- \frac{\partial^2 \Phi_{\rm t}}{\partial x^2} \ \delta x
- \frac{\partial^2 \Phi_{\rm t}}{\partial x \partial y} \delta y
- \frac{\partial^2 \Phi_{\rm t}}{\partial x \partial z} \delta z + \Omega_{\rm b} \delta p_y, \nonumber \\
(\dot{\delta p_y}) &=
- \frac{\partial^2 \Phi_{\rm t}}{\partial y \partial x} \delta x
- \frac{\partial^2 \Phi_{\rm t}}{\partial y^2} \delta y
- \frac{\partial^2 \Phi_{\rm t}}{\partial y \partial z} \delta z - \Omega_{\rm b} \delta p_x, \nonumber \\
(\dot{\delta p_z}) &=
- \frac{\partial^2 \Phi_{\rm t}}{\partial z \partial x} \delta x
- \frac{\partial^2 \Phi_{\rm t}}{\partial z \partial y} \delta y
- \frac{\partial^2 \Phi_{\rm t}}{\partial z^2} \delta z
\label{vareq}
\end{align}

The dynamical system of the barred galaxy has five equilibria known as Lagrange points at which
\begin{equation}
\frac{\partial \Phi_{\rm eff}}{\partial x} = \frac{\partial \Phi_{\rm eff}}{\partial y} = \frac{\partial \Phi_{\rm eff}}{\partial z} = 0.
\label{lgs}
\end{equation}
The collinear points $L_1$, $L_2$, and $L_3$ are located on the $x$-axis, while the central stationary point $L_1$ is a local minimum of $\Phi_{\rm eff}$ at $(x,y,z) = (0,0,0)$. $L_2$ and $L_3$ are saddle points of the effective potential, while $L_4$ and $L_5$ on the other hand are local maxima (see Fig. \ref{isoc}). The stars can move in circular orbits at these four Lagrange points while appearing to be stationary in the rotating frame. The annulus defined by the circles through $L_2$, $L_3$ and $L_4$, $L_5$ is known as the ``region of corotation". The isoline contours of constant effective potential $\Phi_{\rm eff}(x,y,z)$ on the $(x,y)$-plane for $z = 0$ as well as the position of the five Lagrange points $L_i, \ i = {1,5}$ are shown in Fig. \ref{isoc}.

The numerical values of the effective potential at the Lagrange points $L_2$, $L_3$, $L_4$, and $L_5$ are critical values of the Jacobi integral of motion (remember that $E(L_2) = E(L_3)$ and $E(L_4) = E(L_5)$). In particular, for $E > E(L_2)$ the Zero Velocity Curves (ZVCs) are open and two symmetrical channels (exits) are present near the Lagrange points $L_2$ and $L_3$ through which the stars can escape from the interior region of the galaxy. In Fig. \ref{theor}a we present the evolution of the position of the Lagrange points $L_2$ and $L_4$ as a function of the bar's semi-major axis, while in Fig. \ref{theor}b we see the evolution of the critical Jacobi values as a function of $a$. It is evident that as the bar becomes more and more elongated (or in other words stronger) both the position of the Lagrange points and the critical Jacobi values start to diverge.

\section{Orbital dynamics}
\label{orbdyn}

\subsection{Dependence on the semi-major axis of the bar}
\label{semx}

The perturbation scenario as function of $a$ is best presented in the form of a sequence of plots of Poincar\'e maps. As long as we treat bound orbits only the most convenient intersection condition is the one where $R$ runs through a relative maximum and correspondingly the momentum $p_R$ canonically conjugate to the radius $R$ changes from positive to negative values. When we use the radial coordinate to define the intersection, then the natural coordinates for the domain of the map are the canonical coordinates $\phi$ and $L$ of the angular degree of freedom. This map will be called $P$ in the following.

\begin{figure}
\begin{center}
\includegraphics[width=\hsize]{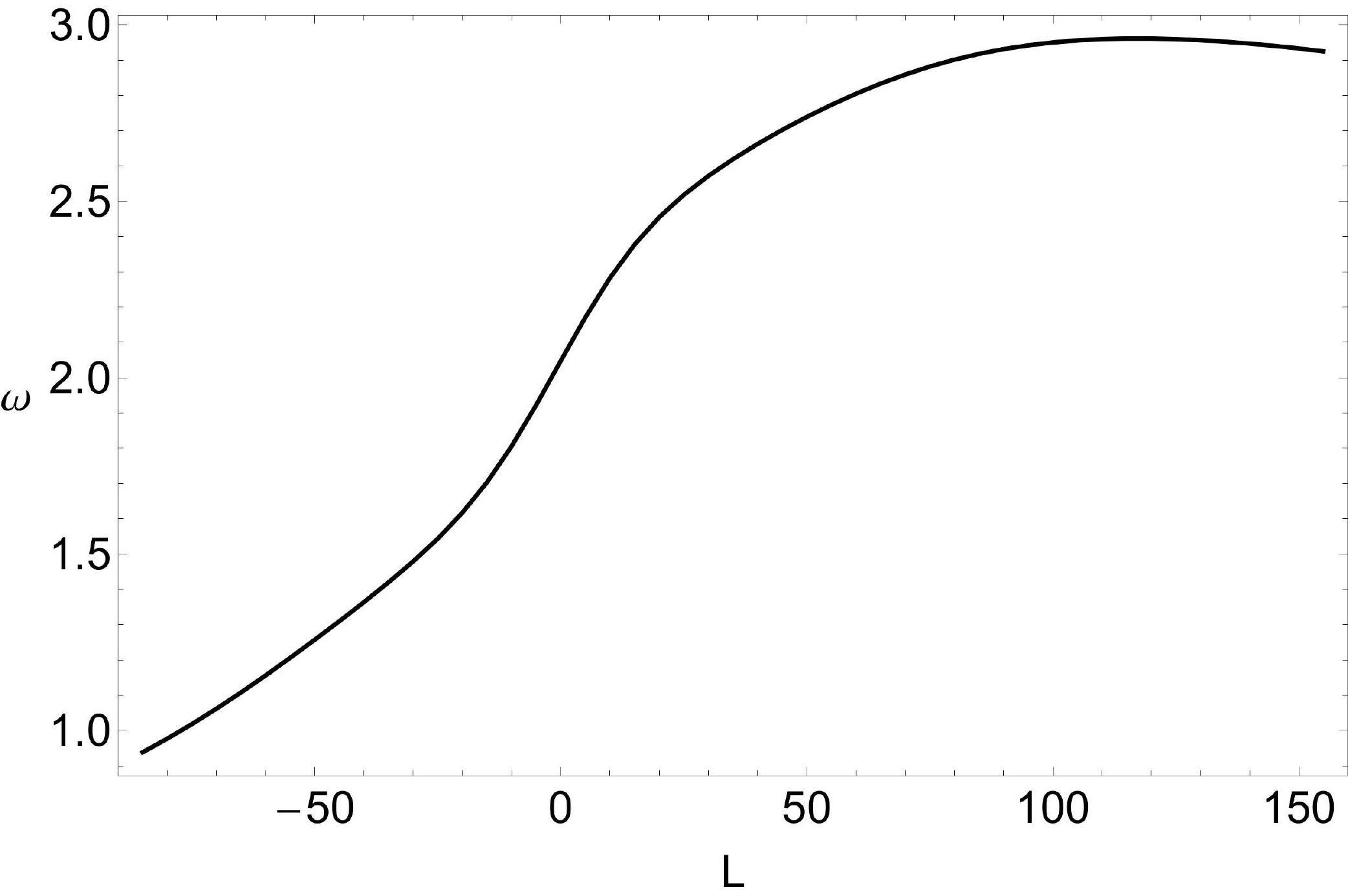}
\end{center}
\caption{The rotation angle $\omega$ as a function of the angular momentum $L$ when $a = 0$.}
\label{tc}
\end{figure}

\begin{figure*}
\centering
\resizebox{0.9\hsize}{!}{\includegraphics{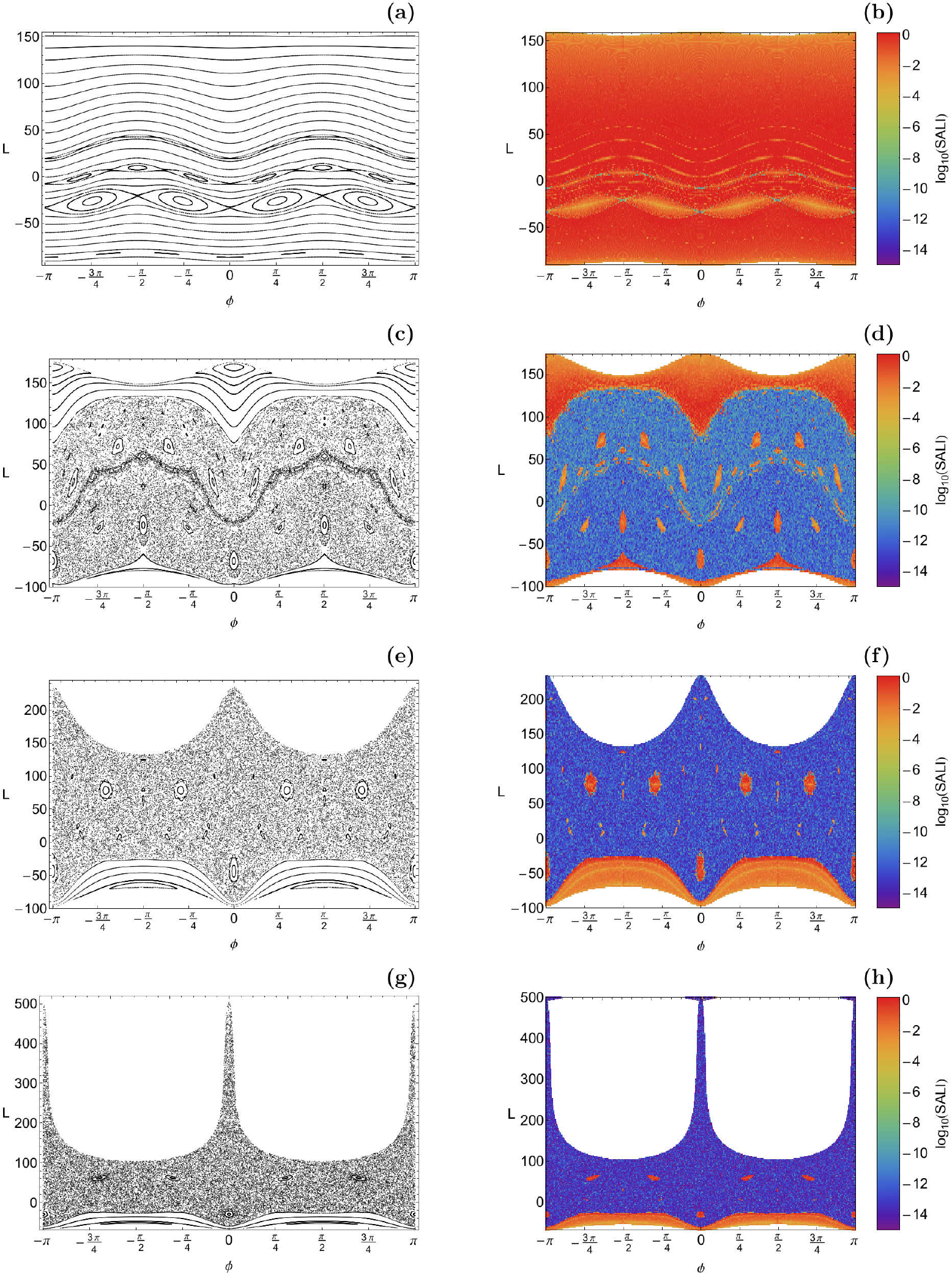}}
\caption{Left column: Examples of the perturbed map $P$ for various values of the semi-major axis of the bar $a$ when $E = -3245$. Right column: Regions of different values of the SALI in the corresponding dense grids of initial conditions on the $(\phi,L)$-plane. Light reddish colors correspond to regular motion, dark blue/purple colors indicate chaotic motion, while all intermediate colors suggest sticky orbits. The energetically forbidden regions of motion are shown in white. (a-b): $a = 1$; (c-d): $a = 2.5$; (e-f): $a = 5$; (g-h): $a = 10$. (\textit{For the interpretation of references to color in this figure caption and the corresponding text, the reader is referred to the electronic version of the article.})}
\label{maps}
\end{figure*}

For $a = 0$ the system is rotationally symmetric, the coordinate $L$ is a conserved quantity and correspondingly the map $P$ is a pure twist map and its domain is foliated into the invariant curves $L = constant$. Under the application of the map each of these invariant curves suffers a rigid shift by a twist angle $\omega$ which is a function of $L$. Fig. \ref{tc} shows this twist curve $\omega(L)$ for the example of $E = -3245$ which is just below $E(L_2)$, i.e. below the escape threshold. To understand the perturbation of the twist map for $a$ differing from 0 it helps to identify the resonances
where $\omega(L)$ runs through rational multiples of $2 \pi$. The most important resonances in our particular case are a 1:6 resonance near $L \approx -80$, a 1:4 resonance near $L \approx -30$, a 1:3 resonance near $L \approx 0$ and a 2:5 resonance near $L \approx 20$. Note that for large values of $L$ between 100 and 150 the twist curve comes close to $\pi$ but does not really reach it. Of course, this twist curve depends on the total orbital energy $E$.

As soon as $a$ is perturbed away from the value 0, the conservation of $L$ is destroyed and the foliation of the domain of $P$ into invariant horizontal lines is lost. For the generic scenario of the perturbation of twist maps see \citet{C79}. Note that the twist curve of Fig. \ref{tc} runs through a local maximum near $L \approx 117$ and then locally around this region the map is not a generic twist map. The left column of Fig. \ref{maps} shows numerical plots of $P$ for the values 1, 2.5, 5 and 10 of $a$. As usual, several initial points have been chosen and many iterates of these initial points are plotted. In the right column of Fig. \ref{maps} we present the corresponding final Smaller ALingment Index (SALI) \citep[The reader can find more information on how this dynamical indicator works in ][]{S01} values obtained from the selected grids of initial conditions, in which each point is colored according to its SALI value at the end of the numerical integration. The value of the SALI indicates the character of an orbit. In particular, after an integration time of $10^4$ time units, we may say that if SALI $> 10^{-4}$ the orbit if regular, while if SALI $< 10^{-8}$ the orbit is chaotic. When $10^{-4} \leq$ SALI $\leq 10^{-8}$ we have the case of a sticky orbit and further numerical integration is needed so as the true nature of the orbit to be fully revealed. It should be pointed out that the SALI method can easily pick out small stability regions embedded in the chaotic sea which can not be easily detected in usual plots of Poincar\'{e} maps. (see e.g., panels d and f of Fig. \ref{maps}). In part (a) of the figure, for $a = 1$, we see a lot of primary invariant curves which are continuous deformations of unperturbed invariant lines. In addition we see some secondary structures, island chains and corresponding fine chaos strips around them which appear very close to separatrix curves. The secondary islands represent resonant coupling between the angular and the radial degree of freedom. The 4 island chains included into the figure correspond exactly to the 4 resonances mentioned above in the discussion of the twist curve.

Part (c) of Fig. \ref{maps} presents the Poincar\'e plot for $a = 2.5$. Here the perturbation is already big enough to destroy the large majority of the primary invariant curves. Only a few ones survive for large positive values of $L$. In addition some secondary islands survive on small scale. The most interesting effect is what happens for large values of $L$ and $\phi$ close to 0 or $\pi$ and for large negative values of $L$ and $\phi$ close to $\pm \pi/2$. First to the events for large values of $L$: We saw in the twist curve that $\omega$ comes close to $\pi$ but does not reach it completely. If it would reach the value $\pi$ then this would imply the existence of corresponding periodic points of period 2 and of an island chain of period 2. However with sufficient perturbation this closeness to the value $\pi$ is already sufficient that the perturbation can bring these periodic points into existence. We see the stable points of period 2 at $\phi = 0$ and $\phi = \pi$ and at $L \approx 170$. In position space they correspond to an orbit oscillating into the long direction of the bar (so called x1 orbit) and when this orbit turns around close to either end of the bar the quantity $R$ runs through a maximum and gives a point for the map at angles 0 and $\pi$. For large negative values of $L$ the anisotropy of the potential for increasing $a$ forces a period 2 orbit into existence which has its longest extension perpendicular to the bar and therefore gives points in the map at angles $\pm \pi/2$. This perpendicular periodic orbit seems to be the most persistent stable periodic orbit of the system. It does not become unstable and does not disappear for any reasonable parameter values.

\begin{figure*}
\centering
\resizebox{\hsize}{!}{\includegraphics{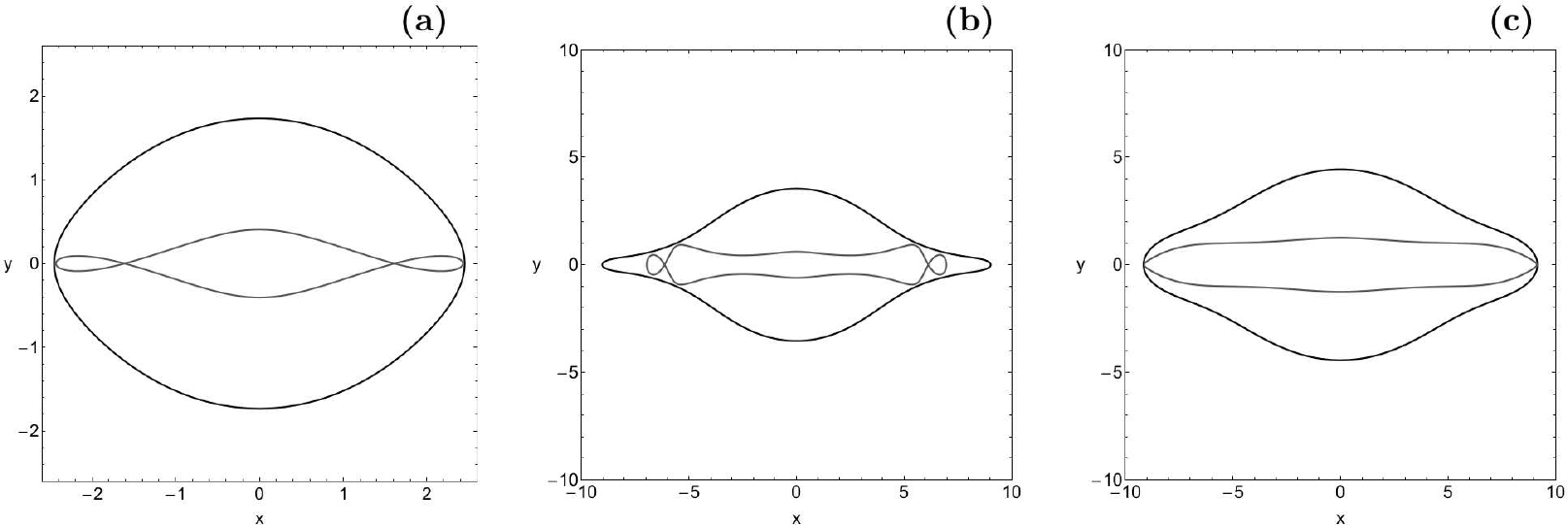}}
\caption{Types of x1 orbits. (a): a 1:3 x1 orbit for $a = 10$ and $E = -4000$; (b): a 1:3 x1 orbit for $a = 10$ and $E = -3300$; (c): a 1:1 x1 orbit for $a = 8.4$ and $E = -3120$. The outermost black solid line is the ZVC.}
\label{x1}
\end{figure*}

For further increasing $a$ the dynamics for large positive values of $L$ becomes unstable, as can be seen in part (e) of Fig. \ref{maps} which presents the map for $a = 5$. The period 2 point at large values of $L$ and at $\phi = 0$ and $\phi = \pi$
becomes unstable near $a \approx 5.4$. It comes back to stability around $a \approx 6.3$ and disappears into the energetic boundary around $a \approx 7.5$. In the large chaotic sea some secondary islands are still visible. Finally in part (g) of Fig. \ref{maps} we present the map for $a = 10$. Near the values 0 and $\pi$ of $\phi$ the domain of the map opens up to very high values of $L$. The explanation is as follows: At energy -3245 we are very close to the energy of the potential saddle of the Lagrange point $L_2$ sitting at $x = R(L_2) = 10.63695596$ and having an energy of $E(L_2) = -3242.77217493$\footnote{In our investigation we used numerical codes with double precision however for simplicity we decided to include in the paper only eight significant decimal figures for all dynamical quantities.}. Therefore for the energy used in the figure orbits can already enter the forming escape channels up to a value of $x$ very close to the Lagrange point. These orbits come near to $L_2$ with a velocity close to 0 in the rotating frame and then they have an angular momentum of approximately $\Omega R(L_2)^2 \approx 500$. These considerations make it understandable that for energies a little higher and a little above $E(L_2)$ the escaping orbits leave the interior region of the effective potential with an angular momentum close to 500.

\subsection{Types of x1 orbits}
\label{x1orb}

An important class of periodic orbits in barred galaxies are the so called x1 orbits which are the backbone of the galactic bar \citep[see e.g.,][]{ABMP83,CP80}. They are orbits surrounding $L_1$ oscillating along the bar within corotation, in our coordinate system it means oscillating in $x$ direction, while in $y$ direction they remain within the width of the bar. We can imagine that the star population of these orbits forms the bar and that these orbits are the building blocks which keep the bar structure dynamically stable. Therefore it is important to describe some details of these orbits in our model.

The existence of this class of orbits at very small energies, i.e. way below the threshold energy $E(L_2)$ is easy to understand. At these extremely small energies the effective potential acts similar to a potential hole around the origin which comes close to a rotationally symmetric one. Then all orbits with small values of the angular momentum oscillate over the origin with a small transverse width. The inclusion of the bar potential breaks the exact rotational symmetry and chooses the $x$ direction and the $y$ direction as normal mode oscillation directions. The x1 orbits are the ones growing out of the $x$ normal mode direction. In part (a) of Fig. \ref{x1} we show a typical example at the energy $E = -4000$, it is for bar length $a = 10$. This type of orbit is called 1:3 resonant x1 orbits since it makes 3 small transverse oscillations in $y$ direction during one oscillation in the longitudinal $x$ direction.

For increasing energy the velocity of the orbit increases, therefore the Coriolis forces increase, and as a consequence the orbit makes wider oscillations in $y$ direction while also the amplitude of the $x$ oscillations increase. Usually the width of the $y$ oscillations grow more rapidly then the amplitude of the $x$ oscillation. In part (b) of Fig. \ref{x1} we show an
example of a 1:3 x1 orbit for the energy $E = -3300$, still below the escape threshold. We already see the tendency to make sharper turns in the $y$ oscillations. For energies above the threshold $E(L_2)$ the orbit forms cusps and extra loops in $y$ direction, its width in $y$ direction becomes larger than 2 and then the orbit no longer qualifies as x1 orbit.

Additional information on x1 orbits is obtained during the study of the scenario of the dynamics under changes of $a$. The x1 orbits run through two maxima of $R$ during a complete period, i.e. a complete oscillation in $x$ direction. Therefore they appear as points of period 2 in the Poincar\'{e} maps presented in the previous section. As seen in the twist curve of Fig. \ref{tc} the twist angle $\omega$ of the unperturbed system, i.e. of the $a = 0$ case does not reach the value $\pi$. However, for large values of $L$ the twist angle $\omega$ comes close to $\pi$. And this has the following consequence: Even though for extremely small values of $a$ we do not have x1 orbits, for mid-size values of $a$ the perturbation brings such orbits into existence. This can be seen best in Fig. \ref{maps}c. The stable elliptic point of period 2 seen at $\phi = 0$ and $\phi = \pi$ and at $L \approx 170$ is the x1 orbit we are looking for.

For moderate values of $a$ it is dynamically stable, for the value $a \approx 5.5$ it becomes dynamically unstable. At the same time with increasing value of $a$ this orbit moves closer to the energetic boundary of the domain of the map and for energy $E = -3245$ it disappears into the energetic boundary at $a \approx 7.5$. Interestingly for energies a little
higher this orbit remains for $a$ values a little higher. For the example $E = -3120$ it disappears for $a \approx 8.5$ only. In part (c) of Fig. \ref{x1} we show this orbit in position space for $a = 8.4$. Along a complete revolution this orbit crosses the $x$-axis once in each of the two orientations and equally crosses also the $y$-axis once in each orientation. In this sense it is a 1:1 x1 orbit.

For high energies around -2000 or -1500 the twist curve crosses the value $\pi$ and then the map $P$ has points of period 2 even at perturbation 0. Unfortunately for such high values of energy the orbits are no longer confined to the interior of the effective potential, they simply leave in any direction and then the concept of x1 orbits no longer makes sense.

We summarize the situation of x1 orbits in our model as follows: They can be found for energies clearly smaller than $E(L_2)$ also for our high values of $\Omega_{\rm b} = 4.5$ and $a = 10$. If $E$ approaches the value $E(L_2)$ with $a = 10$ then a part of them disappears and a part becomes wide in $y$ direction. Acceptable types of x1 orbits only remain for energies above the escape threshold for more moderate values of the bar's semi-major axis.

\section{Escape dynamics}
\label{escdyn}

In this Section we shall try to unveil the escape dynamics of the 2D case $(z = p_z = 0)$ of the SM $(a = 10)$. In particular, our main objective is to determine which orbits escape from the system and which remain trapped. Additionally those orbits which do not escape will be classified using the SALI method into two categories: (i) non-escaping regular orbits and (ii) trapped chaotic orbits. At this point it should be emphasized that in Paper II there was no further classification of non-escaping orbits. Moreover, two important properties of the orbits will be investigated: (i) the exits or channels through which the stars escape and (ii) the time-scale of the escapes (we shall also use the terms escape period or escape rates).

In order to explore the escape dynamics of the model we need to define sets of initial conditions of orbits whose properties will be identified. For this task we define for each value of the Jacobi integral of motion (all tested energy levels are above the escape energy), dense uniform grids of $1024 \times 1024$ initial conditions regularly distributed in the area allowed by the value of the energy. Following a typical approach, all orbits are launched with initial conditions inside the Lagrange radius $(x_0^2 + y_0^2 \leq r_L^2)$, or in other words inside the interior region of the barred galaxy. Our investigation takes place both in the configuration $(x,y)$ and in the phase $(x,p_x)$ space in order to obtain a spherical view of the escape process. Furthermore, the sets of the initial conditions of the orbits are defined as follows: For the configuration $(x,y)$ space we consider orbits with initial conditions $(x_0, y_0)$ with $p_{x_0} = 0$, while the initial value of $p_y$ is always obtained from the Jacobi integral (\ref{ham}) as $p_{y_0} = p_y(x_0,y_0,p_{x_0},E) > 0$. Similarly, for the phase $(x,p_x)$ space we consider orbits with initial conditions $(x_0, p_{x0})$ with $y_0 = 0$, while again the initial value of $p_y$ is obtained from the Jacobi integral (\ref{ham}).

\begin{figure*}
\centering
\resizebox{0.8\hsize}{!}{\includegraphics{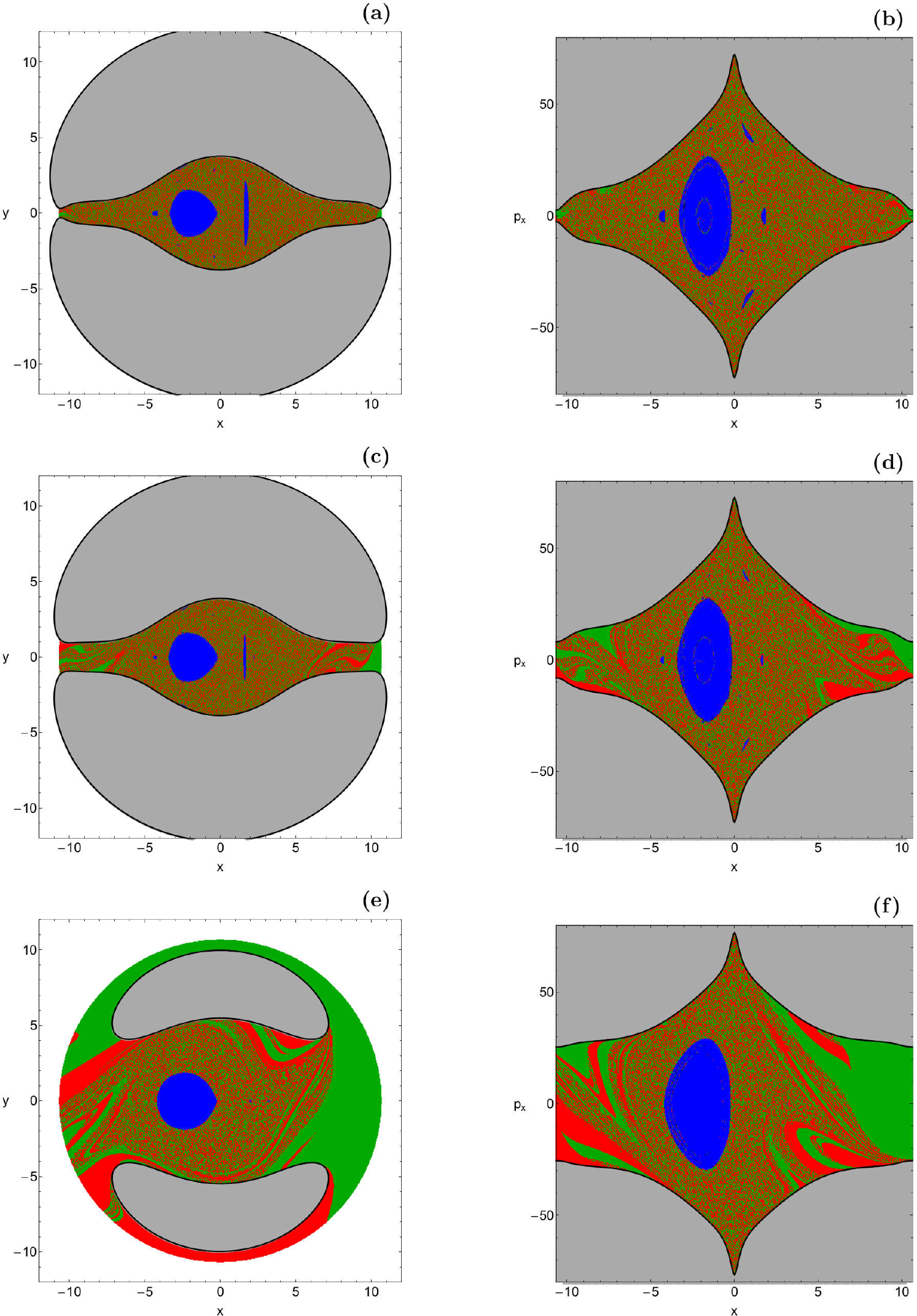}}
\caption{Left column: Orbital structure of the configuration $(x,y)$ space. Right column: Orbital structure of the phase $(x,p_x)$ space. Top row: $\widehat{C} = 0.001$; Middle row: $\widehat{C} = 0.01$; Bottom row: $\widehat{C} = 0.1$. The green regions correspond to initial conditions of orbits where the stars escape through $L_2$, red regions denote initial conditions where the stars escape through $L_3$, blue areas represent stability islands of regular non-escaping orbits, initial conditions of trapped chaotic orbits are marked in black, while the energetically forbidden regions are shown in gray. (\textit{For the interpretation of references to color in this figure caption and the corresponding text, the reader is referred to the electronic version of the article.})}
\label{grd}
\end{figure*}

A double precision Bulirsch-Stoer \verb!FORTRAN 77! algorithm \citep[see e.g.,][]{PTVF92} with a small time step of order of $10^{-2}$, which is sufficient enough for the desired accuracy of our computations was used in order to integrate the equations of motion (\ref{eqmot}) as well as the variational equations (\ref{vareq}) for all the initial conditions of the orbits. Here we would like to emphasize, that our previous numerical experience suggests that the Bulirsch-Stoer integrator is both more accurate and faster than a double precision Runge-Kutta-Fehlberg algorithm of order 7 with Cash-Karp coefficients \citep[see e.g.,][]{DMCG12}. Throughout all our computations, the Jacobi integral of motion (Eq. (\ref{ham})) was conserved better than one part in $10^{-11}$, although for most orbits it was better than one part in $10^{-12}$. All graphics presented in this work have been created using Mathematica$^{\circledR}$ \citep{W03}.

All initial conditions of orbits are numerically integrated for $10^{4}$ time units which correspond to about $10^{12}$ yr. This vast time of numerical integration is justified due to the presence of the sticky orbits\footnote{Sticky orbits are orbits which behave as regular ones during long periods of time before they eventually reveal their true chaotic nature.}. Therefore, if the integration time is too short, any chaos indicator will misclassify sticky orbits as regular ones. In our work we decided to integrate all orbits for a time interval of $10^{4}$ time units in order to correctly classify sticky orbits. At this point, it should be clarified that sticky orbits with sticky periods larger than $10^{4}$ time units will be counted as regular ones, since such extremely high sticky periods are completely unrealistic and of course out of scope of this research.

This critical value of the Jacobi constant at the Lagrange points $L_2$ and $L_3$ $(E(L_2))$ can be used to define a dimensionless energy parameter as
\begin{equation}
\widehat{C} = \frac{E(L_2) - E}{E(L_2)},
\label{chat}
\end{equation}
where $E$ is some other value of the Jacobian. The dimensionless energy parameter $\widehat{C}$ makes the reference to energy levels more convenient. For $\widehat{C} > 0$, the ZVCs are open and therefore stars can escape from the system. The escape dynamics of the system will be investigated for various values of the energy, always within the interval $\widehat{C} \in [0.001,0.1]$.

\begin{figure*}
\centering
\resizebox{\hsize}{!}{\includegraphics{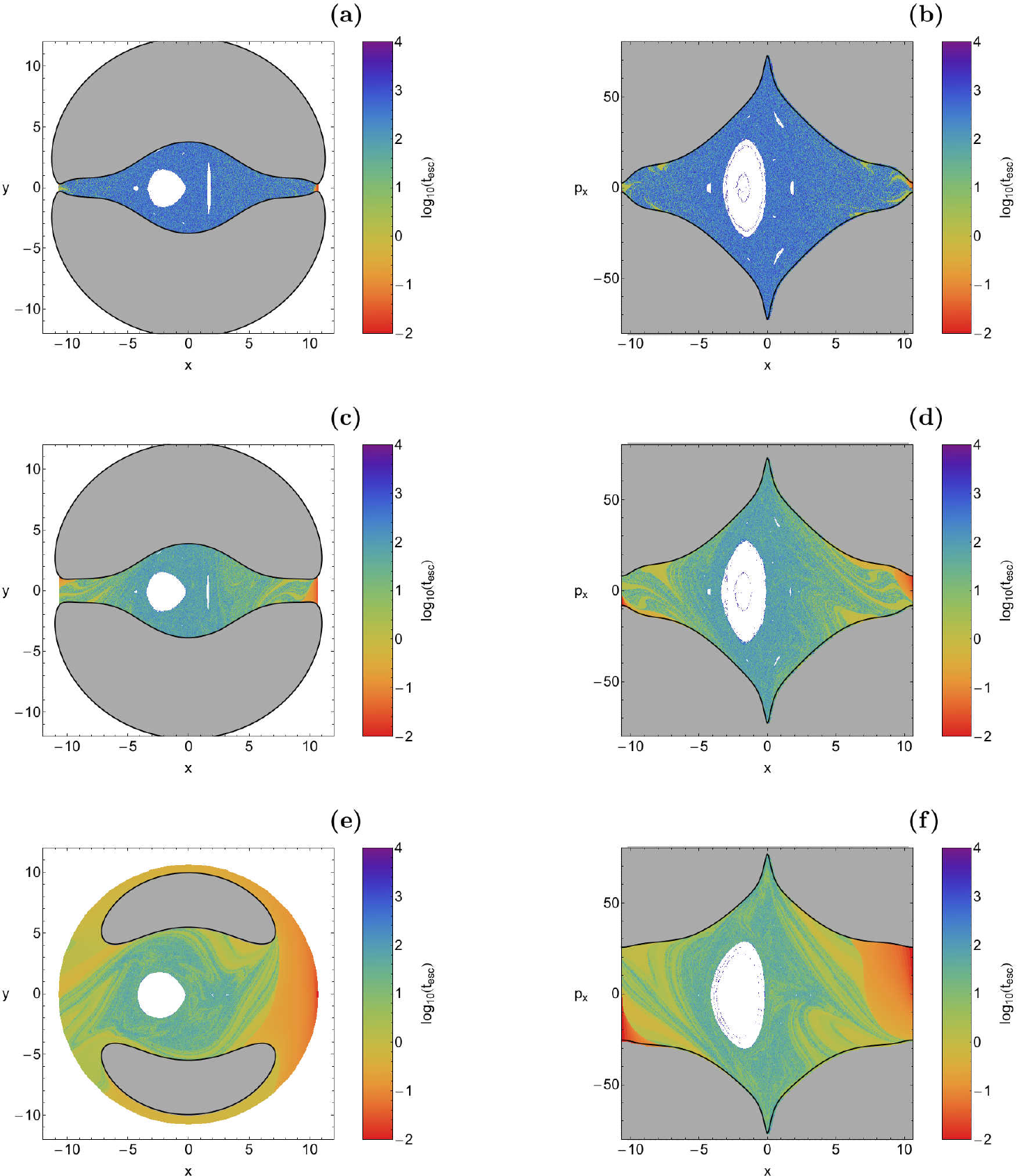}}
\caption{Distribution of the corresponding escape times $t_{\rm esc}$ of the orbits on both types of planes for the corresponding energy levels of Fig. \ref{grd}(a-f). The bluer the color, the larger the escape time. The scale on the color-bar is logarithmic. Initial conditions of non-escaping regular orbits and trapped chaotic orbits are shown in white. (\textit{For the interpretation of references to color in this figure caption and the corresponding text, the reader is referred to the electronic version of the article.})}
\label{times}
\end{figure*}

Usually in an open Hamiltonian system we encounter a mixture of escaping and non-escaping orbits. Generally speaking, the majority of non-escaping orbits are orbits for which a third integral of motion is present, thus restricting their accessible phase space and therefore hinders their escape. However, there are also chaotic orbits which do not escape within the predefined interval of $10^4$ time units and remain trapped for vast periods until they eventually escape to infinity \citep[see e.g.,][]{Z15d}. Here, it should be clarified that these trapped chaotic orbits cannot be considered either as sticky orbits or as super sticky orbits with sticky periods larger than $10^4$ time units. This is true because these trapped orbits exhibit chaoticity very quickly, as it takes no more than about 100 time units for the SALI to cross the threshold value (SALI $\ll 10^{-8}$), thus identifying beyond any doubt their chaotic character. Therefore, we decided to classify the initial conditions of orbits into four main categories: (i) orbits that escape through Lagrange point $L_2$ (right channel or exit 1), (ii) orbits that escape through Lagrange point $L_3$ (left channel or exit 2), (iii) non-escaping regular orbits and (iv) trapped chaotic orbits.

The orbital structure of both the configuration $(x,y)$ and the phase $(x,p_x)$ space for three energy levels is presented in Fig. \ref{grd}(a-f). In these plots the initial conditions of the orbits are classified into four categories by using different colors. Specifically, blue color corresponds to regular non-escaping orbits, black color corresponds to trapped chaotic orbits, green color corresponds to orbits escaping through the Lagrange point $L_2$, while the initial conditions of orbits escaping through the Lagrange point $L_3$ are marked with red color. The outermost black solid line in the configuration space is the ZVC which separates between allowed and forbidden areas (gray color) and it is defined as $\Phi_{\rm eff}(x, y, z = 0) = E$. On the other hand, the outermost black solid line in the phase space is the limiting curve which is defined as
\begin{equation}
f(x,p_x) = \frac{1}{2}p_x^2 + \Phi_{\rm eff}(x, y = 0, z = 0) = E.
\label{zvc}
\end{equation}

The Coriolis forces dictated by the rotation of the galactic bar makes both types of planes asymmetric with respect to the vertical axis and this phenomenon is usually known as ``Coriolis asymmetry" \citep[see e.g.,][]{I80}.
For $\widehat{C} = 0.001$, that is an energy level just above the critical energy of escape $E(L_2)$, we see in Figs. \ref{grd}(a-b) that the vast majority of the initial conditions corresponds to escaping orbits however, a main stability island of non-escaping regular orbits is present denoting bounded ordered motion. It is also seen that the entire escape region is completely fractal\footnote{It should be emphasized that when we state that an area is fractal we simply mean that it has a fractal-like geometry without conducting any specific calculations as in \citet{AVS09}.} which means that there is a strong dependence of the escape mechanism on the particular initial conditions of the orbits. In other words, a minor change in the initial conditions has as a result the star to escape through the opposite escape channel, which is of course, a classical indication of highly chaotic motion. At the outer parts of both planes, near the Lagrange points, we can identify some tiny basins of escape\footnote{An escape basin is defined as a local set of initial conditions of orbits for which the test particles (stars) escape through a certain channel (exit) in the open equipotential surface for energies above the escape energy.}. With a much closer look at the $(x,y)$ and $(x,p_x)$ planes we can identify some additional smaller stability islands which are embedded in the unified escape domain corresponding to secondary resonant orbits. Several basins of escape start to emerge as we proceed to higher energy levels, while the fractal area reduces. Indeed for $\widehat{C} = 0.01$ we observe in Fig. \ref{grd}(c-d) that at the outer parts of both types of planes several well-defined basins of escape are present. In addition, the extent of the main 1:1 stability island seems to be unaffected by the increase in the orbital energy, while on the other hand the area occupied by secondary resonances seems to decrease. At the highest value of the energy studied, that is $\widehat{C} = 0.1$, one may observe in Fig. \ref{grd}(e-f) that broad well-formed basins of escape dominate both types of planes, while the fractal regions are confined mainly near the boundaries between the escape basins or in the vicinity of the stability islands. Furthermore, all the stability regions corresponding to secondary resonances disappear and only the main 1:1 resonance survives. It should be noted that in the $(x,p_x)$ phase space inside the main 1:1 stability island we can distinguish the existence of thin structures (green and red rings) composed of initial conditions of escaping orbits. These initial conditions correspond to chaotic orbits which form thin chaotic rings inside the stability region.

Our numerical computations suggest that trapped chaotic motion in the 2D case of the barred galaxy model is almost negligible. This is true because the initial conditions of such orbits appear only as lonely points around the boundaries of the stability islands. In all studied energy levels the areas of regular motion correspond mainly to retrograde orbits (i.e., when a star revolves around the galaxy in the opposite sense with respect to the motion of the galaxy itself), while there are also some smaller stability islands of prograde orbits.

The escape dynamics of our barred galaxy model presented in Figs. \ref{grd}(a-f) is very different with respect to that observed in the barred galaxy model investigated in Paper II (see Fig. 5). In particular, in Paper II it was found that the vast majority of both the configuration and the phase space corresponds to non-escaping orbits (both regular and chaotic). In our dynamical model on the other hand, non-escaping regular orbits occupy only a small fraction of the grids (less than 50\%), while the remaining area is covered either by fractal domains or by basins of escape. Therefore we may conclude that the present model contains a much more interesting and realistic escape dynamics than that of Paper II.

\begin{figure*}
\centering
\resizebox{\hsize}{!}{\includegraphics{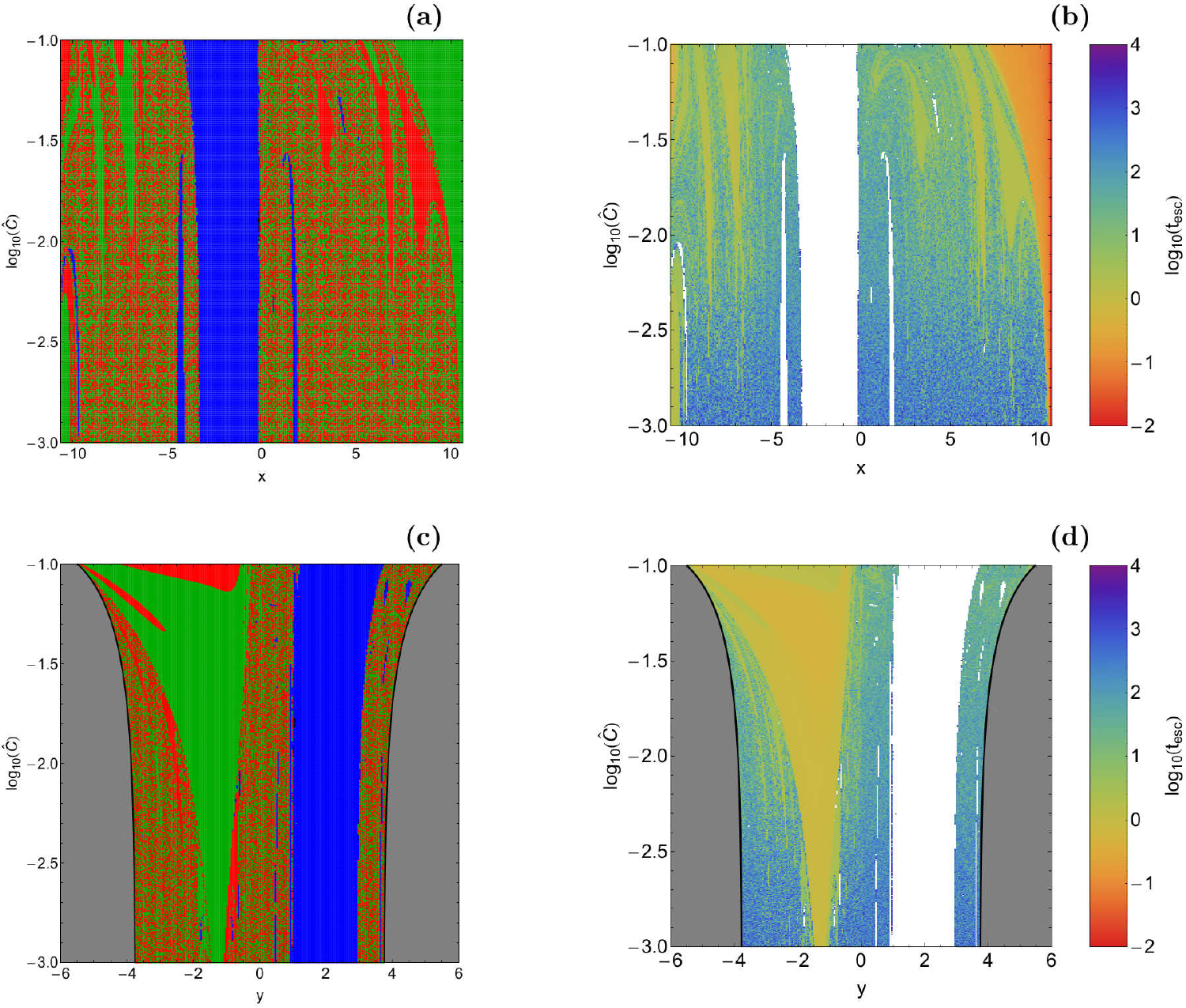}}
\caption{Orbital structure of the (a-upper left): $(x,\widehat{C})$-plane; and (c-lower left): $(y,\widehat{C})$-plane; (b-upper right and d-lower right): the distribution of the corresponding escape times of the orbits. The color code is exactly the same as in Fig. \ref{grd}. (\textit{For the interpretation of references to color in this figure caption and the corresponding text, the reader is referred to the electronic version of the article.})}
\label{xytc}
\end{figure*}

\begin{figure*}
\centering
\resizebox{\hsize}{!}{\includegraphics{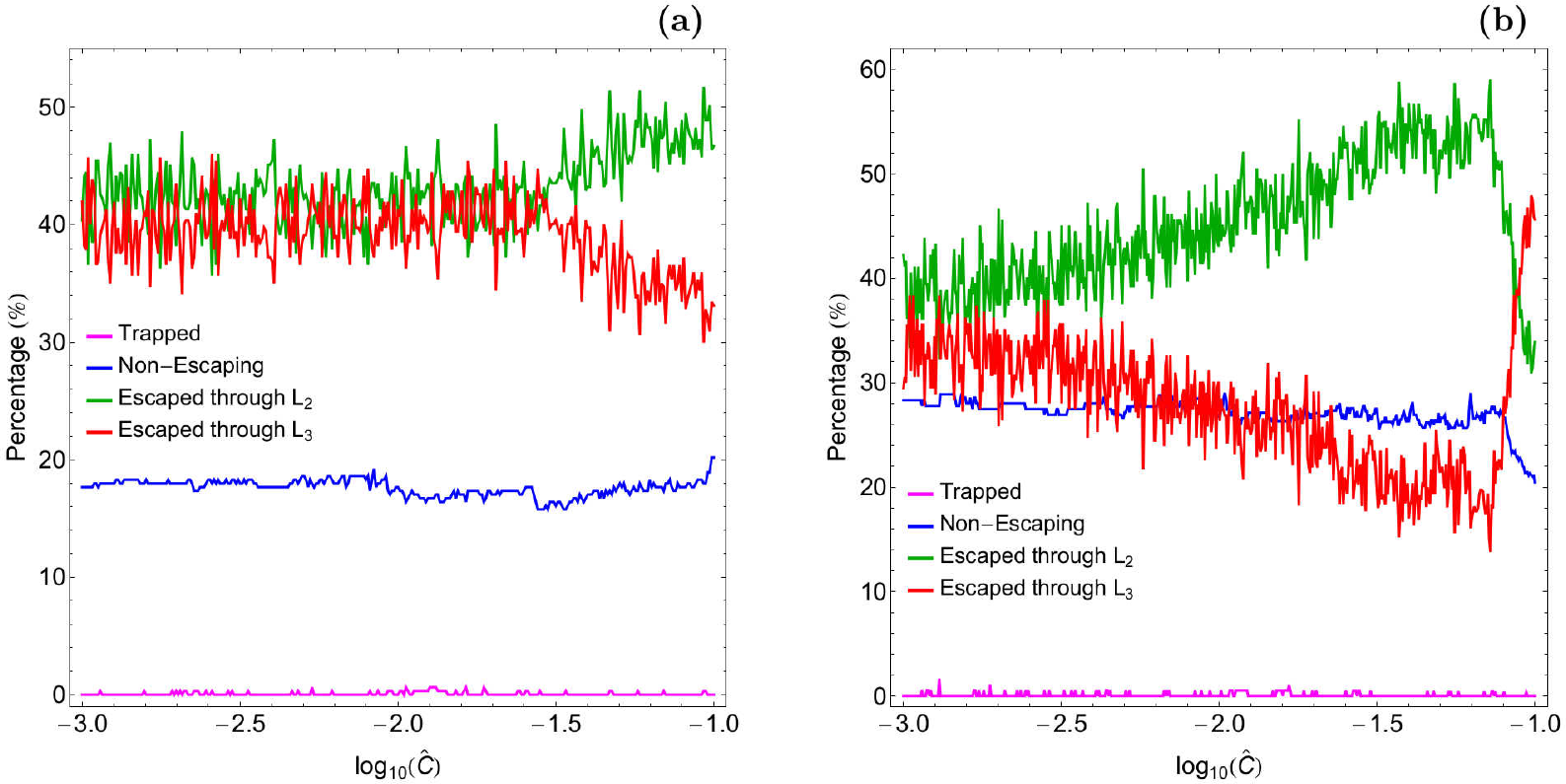}}
\caption{Evolution of the percentages of escaping, non-escaping regular and trapped chaotic orbits on the (a-left): $(x,\widehat{C})$-plane and (b-right): $(y,\widehat{C})$-plane as a function of the dimensionless energy parameter $\widehat{C}$. (\textit{For the interpretation of references to color in this figure caption and the corresponding text, the reader is referred to the electronic version of the article.})}
\label{percs}
\end{figure*}

The distribution of the escape times $t_{\rm esc}$ of orbits on both types of planes is presented in Fig. \ref{times}(a-f), where the energy levels are exactly the same as in Fig. \ref{grd}(a-f). Initial conditions of fast escaping orbits, with short escape times, correspond to light reddish colors, dark blue/purple colors indicate large escape times, while regular non-escaping and trapped chaotic orbits are shown in white. Note that the scale on the color bar is logarithmic. As expected, orbits with initial conditions near the vicinity of the stability islands or inside the fractal regions need significant amount of time in order to escape from the barred galaxy, while on the other hand, inside the basins of escape, where there is no dependence on the initial conditions whatsoever, we measured the shortest escape rates of the orbits. For $\widehat{C} = 0.001$ we observe that the escape times of orbits with initial conditions inside the fractal region of the planes are huge corresponding to tens of thousands of time units. However as we proceed to higher values of the energy the escape times of the orbits significantly reduce. The basins of escape can also be distinguished in Fig. \ref{times}(a-f) being the regions with reddish colors indicating extremely fast escaping orbits. Indeed our numerical calculations indicate that orbits with initial conditions inside the basins of escape have significantly small escape times of less than 10 time units. Our numerical analysis indicate that the average escaping time of the orbits which form, in the phase space, the escape rings inside the main 1:1 stability region is more than $10^3$ time units.

We may conclude that as the value of the energy increases three interesting phenomena take place in both types of planes: (i) the regions of forbidden motion are confined (especially in the configuration space), (ii) the two escape channels become wider and wider and (iii) the escape times of orbits are reduced.

The color-coded plots (see Fig. \ref{grd}(a-f)) in the configuration $(x,y)$ as well as in the phase $(x,p_x)$ space provide sufficient information on the phase space mixing however for only a fixed value of the Jacobi constant $E$. However H\'{e}non \citep{H69}, in order to overcome this drawback introduced a new type of plane which can provide information not only about stability and chaotic regions but also about areas of trapped and escaping orbits using the section $y = p_x = 0$, $p_y > 0$ \citep[see also][]{BBS08}. In other words, all the orbits of the stars of the barred galaxy are launched from the $x$-axis with $x = x_0$, parallel to the $y$-axis $(y = 0)$. Consequently, in contrast to the previously discussed types of planes, only orbits with pericenters on the $x$-axis are included and therefore, the value of the dimensionless energy parameter $\widehat{C}$ can be used as an ordinate. In this way, we can monitor how the energy influences the overall orbital structure of our dynamical system using a continuous spectrum of energy values rather than few discrete energy levels. We decided to investigate the energy range when $\widehat{C} \in [0.001,0.1]$.

The orbital structure of the $(x,\widehat{C})$-plane when $\widehat{C} \in [0.001,0.1]$ is presented in Fig. \ref{xytc}a, while in Fig. \ref{xytc}b the distribution of the corresponding escape times are given. It is seen that for relatively low values of the energy $(0.001 < \widehat{C} < 0.01)$ the central region of the $(x,\widehat{C})$-plane is highly fractal, while only at the outer parts of the same plane we can identify some weak basins of escape. The main stability island corresponding to retrograde $(x_0 < 0)$ non-escaping regular 1:1 orbits remains almost unperturbed by the increase of the total orbital energy. Two smaller stability islands of secondary resonances are visible in the negative part of the $(x,\widehat{C})$-plane, while a prograde $(x_0 > 0)$ stability island is also present. As the value of the energy increases the structure of the $(x,\widehat{C})$-plane changes drastically and the most important differences are the following: (i) all stability islands corresponding to secondary resonances disappear, while only the main 1:1 resonance survives; (ii) several basins of escape emerge, while the fractal regions are confined to the boundaries of the escape basins and in the vicinity of the stability island; (iii) the area covered by the escape basin at the outer right part of the $(x,\widehat{C})$-plane increases meaning that at high enough energy levels the escape through Lagrange point $L_2$ is more preferable.

In an attempt to obtain a more complete and spherical view about the escape dynamics in our barred galaxy model we follow a similar numerical approach to that explained earlier but this time all orbits of the stars are initiated from the vertical $y$-axis with $y = y_0$. In particular, we use the section $x = p_y = 0$, $p_x > 0$, launching orbits parallel to the $x$-axis. This technique allow us to construct again a two-dimensional (2D) plane in which the $y$ coordinate of orbits is the abscissa, while the logarithmic value of the energy $\log_{10}(\widehat{C})$ is the ordinate. The orbital structure of the $(y,\widehat{C})$-plane when $\widehat{C} \in [0.001,0.1]$ is shown in Fig. \ref{xytc}c. The outermost black solid line is the limiting curve which distinguishes between regions of forbidden and allowed motion and is defined as
\begin{equation}
f_L(y,\widehat{C}) = \Phi_{\rm eff}(x = 0, y, z = 0) = E.
\label{zvc2}
\end{equation}
A very complicated orbital structure is unveiled in the $(y,\widehat{C})$-plane. This structure has two main differences with respect to that discussed previously for the $(x,\widehat{C})$-plane. Being more precise: (i) the main 1:1 stability island is now located in the $y > 0$ part of the plane and (ii) a basin of escape corresponding to escape through exit 1 is always present in the $y < 0$ part of the plane. Furthermore, the extent of this basin of escape grows with increasing energy and for $\widehat{C} = 0.1$ it occupies about half of the $(x,\widehat{C})$-plane. However, it should be noted that for high enough values of the energy a smaller basin of escape corresponding to exit 2 emerges inside the basin of exit 1.

Looking at Figs. \ref{xytc}b and \ref{xytc}d we observe that the escape times of the orbits are strongly correlated with the escape basins. We may conclude that the smallest escape times correspond to orbits with initial conditions inside the escape basins, while orbits with initial conditions in the fractal regions of the planes or near the boundaries of stability islands have the highest escape rates. In both types of planes the escape times of orbits are significantly reduced with increasing energy. The connection with the invariant manifolds of the outermost periodic orbits which define the basin boundaries and direct the outgoing flow will be explained in the next Section.

\begin{figure*}
\centering
\resizebox{\hsize}{!}{\includegraphics{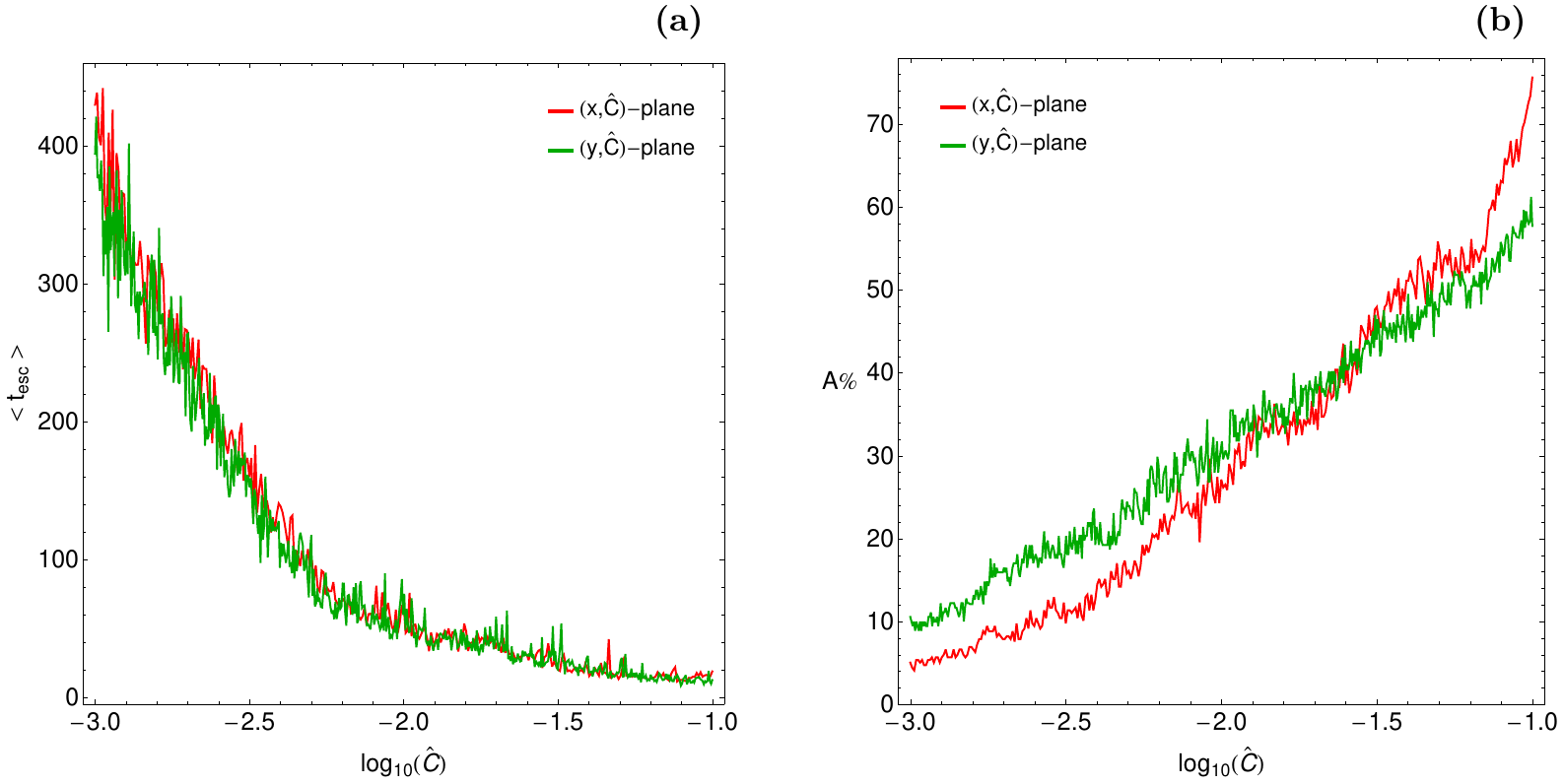}}
\caption{(a-left): The average escape time of orbits $< t_{\rm esc} >$ and (b-right): the percentage of the total area $A$ of the planes covered by the escape basins as a function of the dimensionless energy parameter $\widehat{C}$. (\textit{For the interpretation of references to color in this figure caption and the corresponding text, the reader is referred to the electronic version of the article.})}
\label{stats}
\end{figure*}

The evolution of the percentages of the four types of orbits on the $(x,\widehat{C})$ and $(y,\widehat{C})$ planes as a function of the dimensionless energy parameter $\widehat{C}$ is shown in Fig. \ref{percs}(a-b), respectively. It is seen in Fig. \ref{percs}a that for about $\widehat{C} < 0.03$ the percentages of escaping orbits through exits 1 and 2 display similar fluctuations, while for higher values of the energy their rates start to diverge. In particular, the percentages of escaping orbits through the Lagrange point $L_2$ increases up to about 50\%, while the rate of escapers through $L_3$ decreases up to about 30\%. The percentage of non-escaping regular orbits remain, in general terms, almost unperturbed around 16\% to 20\%, while that of trapped chaotic orbits is zero in most of the cases apart from some non-zero peaks. In the same vein we observe in Fig. \ref{percs}b the evolution of the percentages for the $(y,\widehat{C})$ plane. Here the percentages of escaping orbits diverge much sooner at about $\widehat{C} = 0.002$. The rate of escapers through exit 1 increases, while that of escapers through exit 2 decreases, although that for high enough values of the energy $(\widehat{C} > 0.07)$ this trend is reversed. The amount of non-escaping regular orbits remain again almost unperturbed at about 28\% but for $\widehat{C} > 0.07$ their percentage exhibit a minor decrease. Once more the rate of trapped chaotic orbits is almost negligible.

Finally in Fig. \ref{stats}a we present the evolution of the average value of the escape time $< t_{\rm esc} >$ of orbits as a function of the dimensionless energy parameter for the $(x,\widehat{C})$ and $(y,\widehat{C})$ planes. We observe that for low energy levels, just above the critical energy of escape, the average escape period of orbits is more than 400 time units. However as the value of the energy increases the escape time of the orbits reduces rapidly tending asymptotically to zero which refers to orbits that escape almost immediately from the system. If we want to justify the behaviour of the escape time we should take into account the geometry of the open ZVC. In particular, as the total orbital energy increases the two symmetrical escape channels near the saddle points become more and more wide and therefore, the stars need less and less time until they find one of the two openings (holes) in the ZVC and escape from the system. This geometrical feature explains why for low values of the Jacobi integral orbits consume large time periods wandering inside the open ZVC until they eventually locate one of the two exits and escape from the system. The evolution of the percentage of the total area $(A)$ on the $(x,\widehat{C})$ and $(y,\widehat{C})$ planes corresponding to basins of escape, as a function of the dimensionless energy parameter is given in Fig. \ref{stats}b. As expected, for low values of the total orbital energy the degree of fractality on both types of planes is high. However, as the energy increases the rate of fractal domains reduces and the percentage of domains covered by basins of escape starts to grow rapidly. Eventually, at relatively high energy levels $(\widehat{C} = 0.1)$ the fractal domains are significantly confined and therefore the well formed basins of escape occupy more than 70\% of the $(x,\widehat{C})$-plane and more than 60\% of the $(y,\widehat{C})$-plane.

\section{Lyapunov orbits and invariant manifolds}
\label{loman}

\begin{figure*}
\centering
\resizebox{\hsize}{!}{\includegraphics{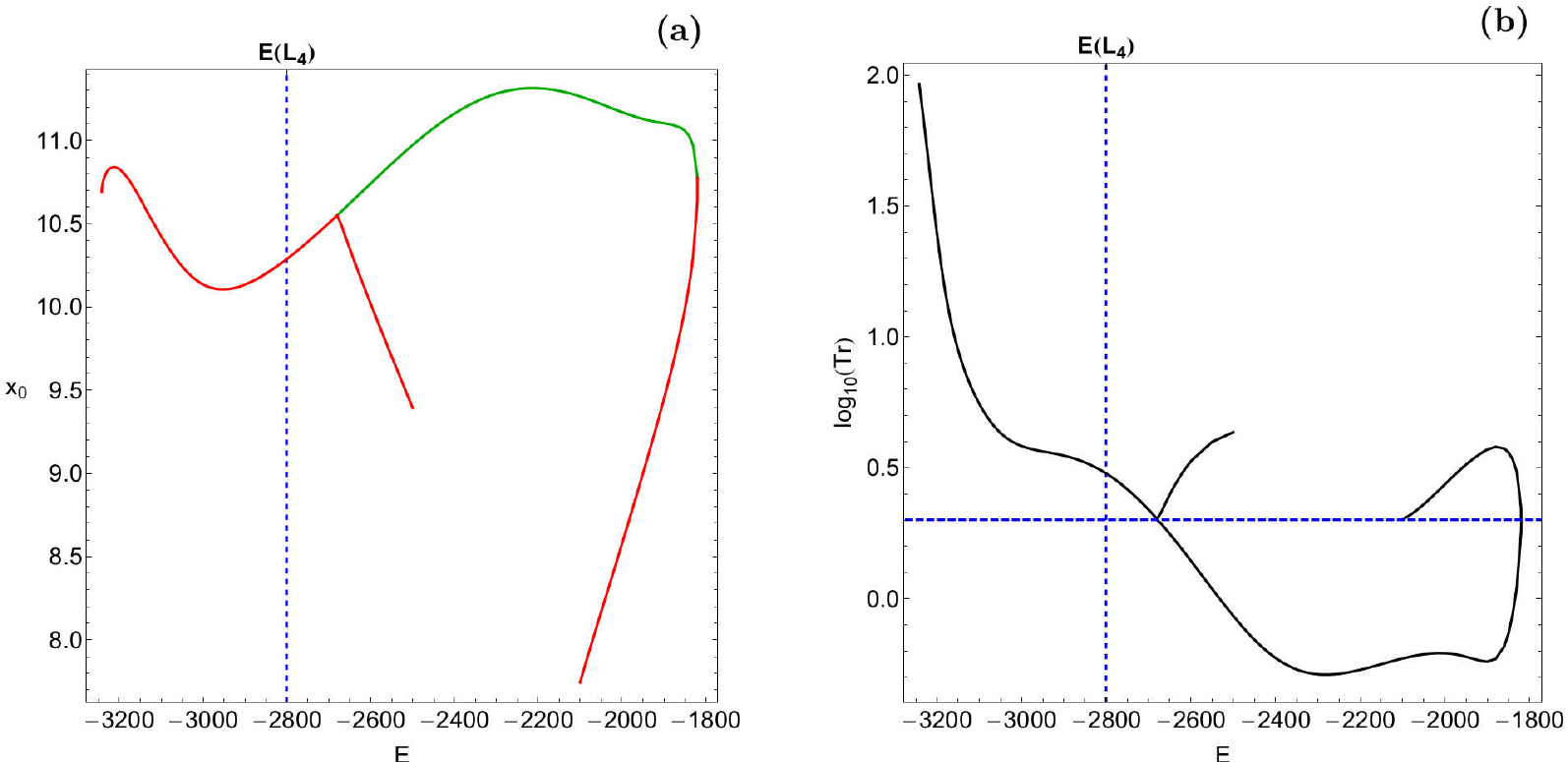}}
\caption{(a-left): Evolution of the $x$ coordinate of the outermost point along $LO_2$ as a function of the energy. Green color indicates initial conditions of stable periodic orbits, while initial conditions of unstable periodic orbits are shown in red. (b-right): The trace of the monodromy matrix of $LO_2$, as a function of the energy $E$. The horizontal blue dashed line corresponds to the critical value Tr = +2, which distinguishes between stable and unstable motion. (\textit{For the interpretation of references to color in this figure caption and the corresponding text, the reader is referred to the electronic version of the article.})}
\label{los}
\end{figure*}

The Lagrange points $L_2$ and $L_3$ are the outermost saddle points of the effective potential from Eq. (\ref{Veff}). Therefore they are linked to the outermost elements of the chaotic invariant set, in this case periodic orbits which are normal hyperbolic at least for energies close to the saddle energy. These orbits are also called in plane Lyapunov orbits \citep{L49} and have a roughly elliptical shape. In the following we denote them as $LO_2$ and $LO_3$, respectively. They exist for energy $E$ larger then the saddle energy $E(L_2)$ and persist up to an upper energy limit mentioned below. Because of discrete symmetry of the system one of these orbits is obtained from the other one by a rotation of the whole system by $\pi$ around the origin. Therefore it is sufficient to give details for the one over the saddle point $L_2$ only.

Part (a) of Fig. \ref{los} shows as function of the energy the $x$ coordinate of the outermost point along $LO_2$ and this point is at the same time the one which fulfils the intersection condition of the Poincar\'e map. Because of symmetry reasons it lies at $\phi = 0$. The original Lyapunov orbit $LO_2$ is the upper branch of the curves shown. The lower branch of the curves in part (b) of the same figure shows the trace of the monodromy matrix of $LO_2$. The orbit is in plane unstable as
long as this trace is larger than 2 and stable when its trace is between -2 and +2. Because of the importance of the value 2 it is marked in the figure as dashed horizontal line (blue in the colour version).

\begin{figure}
\begin{center}
\includegraphics[width=\hsize]{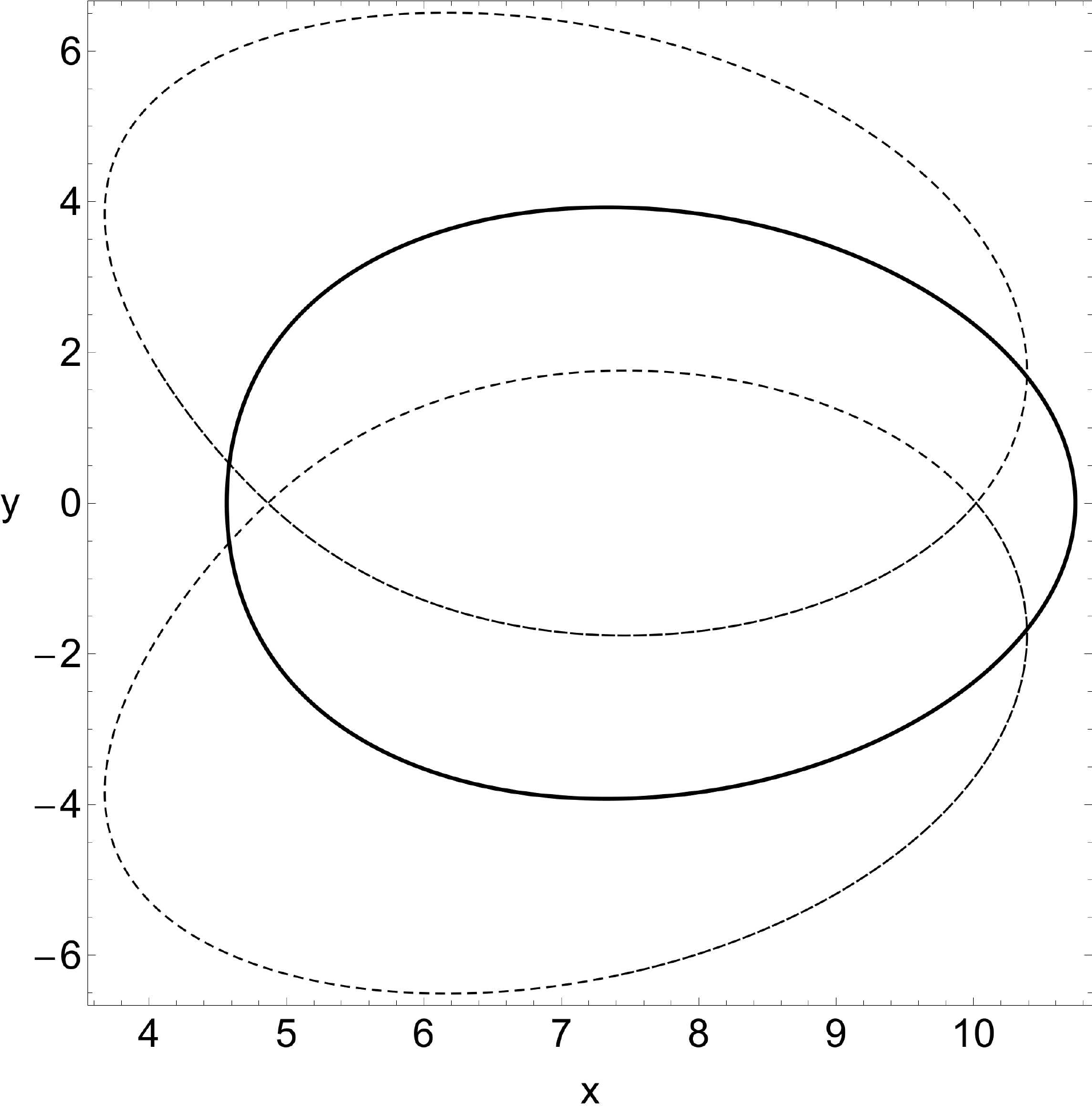}
\end{center}
\caption{The $LO_2$ periodic orbit (solid line) and its two descendants (dashed lines) for $E = -2600$ and $a = 10$.}
\label{pos}
\end{figure}

\begin{figure*}
\centering
\resizebox{\hsize}{!}{\includegraphics{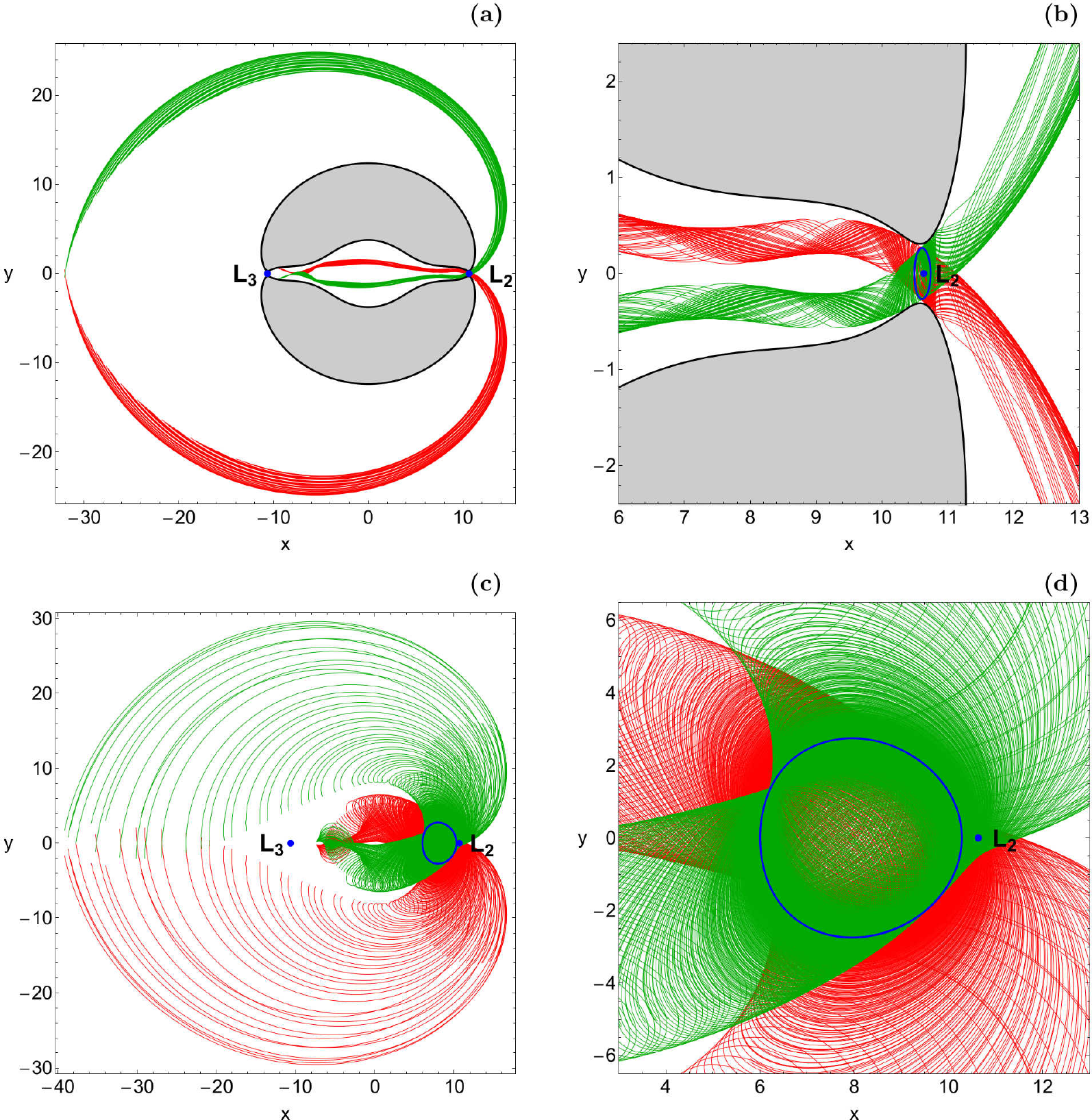}}
\caption{The stable manifold $W^s(LO_2)$ (green), the unstable manifold $W^u(LO_2)$ (red), and the corresponding Lyapunov orbit (blue) when $a = 10$, for (a-b): $\widehat{C} = 0.001$ and (c-d): $E(L_4)$ ($\widehat{C} = 0.13638598$). (\textit{For the interpretation of references to color in this figure caption and the corresponding text, the reader is referred to the electronic version of the article.})}
\label{mans}
\end{figure*}

\begin{figure*}
\centering
\resizebox{\hsize}{!}{\includegraphics{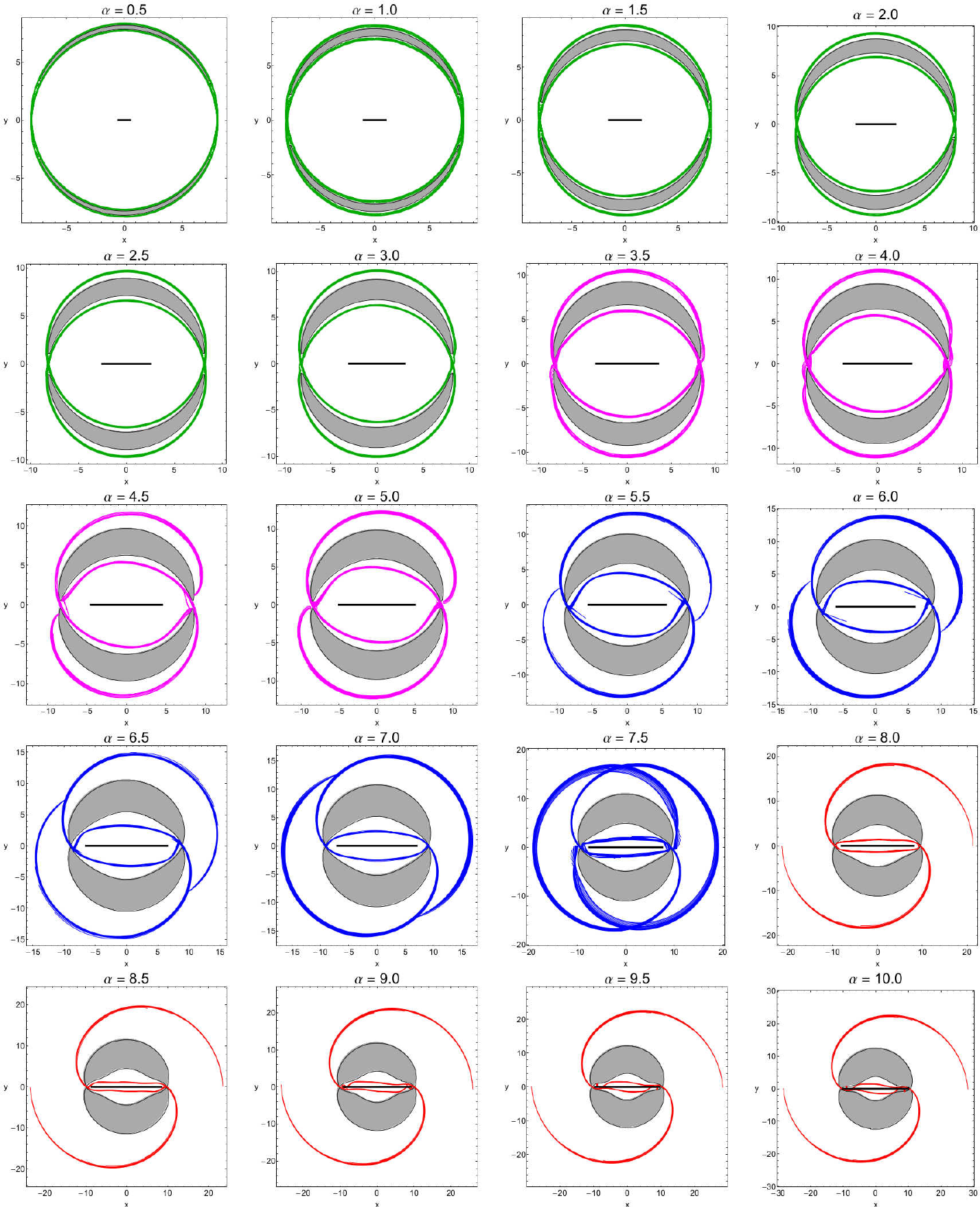}}
\caption{Morphologies of the unstable manifolds $W^u(LO_2)$ and $W^u(LO_3)$ for several values of the bar's semi-major axis $a$, while for all models $\widehat{C} = 0.0001$. The manifolds are plotted in a different colour which is determined by their morphology: $R_1$ rings (green); $R_1'$ pseudo-rings (magenta); $R_1R_2$ rings (blue); open spirals (red). The horizontal black lines in the interior region indicate the total length of the bar. (\textit{For the interpretation of references to color in this figure caption and the corresponding text, the reader is referred to the electronic version of the article.})}
\label{col}
\end{figure*}

We observe that at $E(L_2)$ the orbit $LO_2$ is born unstable, its instability however becomes smaller with increasing energy and at $E_p \approx -2680$ it becomes stable in a pitchfork bifurcation when it splits off two unstable descendants. This behavior (the Lyapunov orbits to become stable for energies higher than $E(L_2)$) is consistent with previous studies using different dynamical models \citep[see e.g.,][]{SPA02}. Each one of these descendants lies asymmetrically to the $x$-axis, but one of them is obtained from the other one by a reflection in the $x$-axis. Therefore the two descendants have the same stability properties and the same intersection coordinate with the $x$-axis. The stability trace of the two descendants is included into part (b) of Fig. \ref{los} as the curve which branches off from the main curve at $E_p$. In part (a) of the same figure we also included the intersection coordinate of the descendants with the $x$-axis as the curve which branches off from the main curve at $E_p$. $LO_2$ disappears in a saddle center bifurcation at $E \approx -1820$. Here it collides with an unstable periodic orbit coming from the interior of the potential well. In Fig. \ref{los} also the data of this periodic orbit are included. In the colour version of part (a) of the same figure red means an unstable orbit and green means a stable one. Fig. \ref{los} is plotted for the parameter value $a = 10$. Fig. \ref{pos} shows in the position space for $E = -2600$ and for $a = 10$ the orbit $LO_2$ and its two descendants created in the pitchfork bifurcation.

For energies in the interval $I = [E(L_2), E_p]$ the stable and unstable manifolds of $LO_2$ direct the flow over the saddle $L_2$, i.e. the flow from the interior of the potential well to the exterior and in opposite direction as well. Therefore we have to understand these invariant manifolds quite well and we want to relate them to structures observed in the outer parts of the galaxy. In the following we call these stable and unstable manifolds $W^s(LO_2)$ and $W^u(LO_2)$, respectively. Of course, these manifolds live in the energy shell of the phase space. In order to present figures we project them into the position $(x,y)$ space. $LO_2$ itself has the topology of a circle and the local branches of $W^s(LO_2)$ and $W^u(LO_2)$ have the topology of tubes. The local segments of these two tubes intersect exactly in the periodic orbit itself. If we continue the stable and unstable manifolds beyond their local segments then globally they form an infinity of tendrils which contain an infinity of further mutual intersections, the so called homoclinic points. In the present article we do not study these global properties nor the homoclinic structure, we concentrate on the local structure only. The reader can find more information about the invariant manifolds in \citet{RGMA06} and \citet{RGAM07}.

Numerically and graphically we obtain an impression of the local segments of $W^u(LO_2)$ by choosing a set of initial conditions in the vicinity of $LO_2$ and by following the resulting orbits. These orbits are trapped inside the manifolds, so that they stay confined together, at least for a couple of rotations around the galactic bar. This provides a set of curves on $W^u(LO_2)$ which give a rather good optical impression of the tubular surface itself. The corresponding representation of $W^s(LO_2)$ is obtained by letting the same orbits run backwards in time. In the following numerical plots $W^u(LO_2)$ is drawn by red lines and $W^s(LO_2)$ is represented by green lines, while the Lyapunov orbit $LO_2$ is drawn blue. At this point it should be pointed out that the dynamics of the manifolds do not affect all the orbits, only those with initial conditions near the in plane Lyapunov periodic orbits.

Part (a) of Fig. \ref{mans} presents the resulting structure for $\widehat{C} = 0.001$ and $a = 10$, while part (b) is a magnification of part (a) and shows in better resolution the neighbourhood of the saddle point $L_2$. We observe quite well
how individual orbits wind around the tubes of the invariant manifolds. The black curve in the same figure is the ZVC and its interior (gray shaded in the plot) is the energetically forbidden region of the position space. The saddle points $L_2$ and $L_3$ of the effective potential are marked as blue dots. Parts (c) and (d) of the Fig. \ref{mans} show the corresponding structures at the higher energy $E(L_4) \approx -2800$ ($\widehat{C} = 0.13638598$) where the whole position space becomes energetically accessible.

The interest in the invariant manifolds of the Lyapunov orbits comes from the following considerations: Stars which leave the interior region of the effective potential do so close to the unstable manifolds of the outermost elements of the chaotic invariant set and the orbits in position space mark regions of perturbation of the disk structure. These perturbations
may trigger star formation and cause the formation of spirals and rings and keep them stable. We can think of the manifolds as channels through which material can be transported between different parts of the galaxy such as between the regions inside and outside the corotation \citep[see e.g.,][]{GKL04,KLMR00}. Therefore we expect a direct connection between the projection of the local segments of $W^u(LO_2)$ and $W^u(LO_3)$ into the position space on one hand and the spirals and rings of barred galaxies on the other hand. These structures are closely related to the structure of the bar and therefore should depend strongly on the numerical value of the bar's semi-major axis $a$. In Fig. \ref{col} we present the local segments of $W^u(LO_2)$ and $W^u(LO_3)$ for values of $a$ from 0.5 up to 10 in steps of 0.5. The energy of each part is chosen such that it is $\widehat{C} = 0.0001$ above the respective saddle energy which also depends on the bar's semi-major axis. Here it should be pointed out that the existence of the unstable manifolds is only a necessary but not at all a sufficient condition for the corresponding structure to develop. In addition, theoretically (using the numerical integration) a manifold may be present in a dynamical model. However in a real galaxy with similar dynamical properties (like those of the mathematical model) the manifold may not be able to trap sufficient orbits inside it, so the corresponding stellar structure (ring or spiral) will not be observable.

At first sight a clear development scenario is obvious. For small values $a \in [0.5,3.0]$ the local segments of the unstable manifolds trace out a ring structure around the interior potential region with the longer axis of the ring in $y$ direction. This structure is called $R_1$ rings. Here the unstable manifolds from one side come very close to the opposite saddle point of the potential forming approximate heteroclinic separatrix connections. For this situation the manifolds are plotted green in the figures. For $a \in [3.5,5.0]$ the heteroclinic connections are clearly broken and in addition the major axis of the ring rotates in negative orientation with increasing $a$. The resulting structures are called $R_1'$ pseudo-rings, and they are plotted in magenta in the figures. For $a \in [5.5,7.5]$ the unstable manifolds coming from one side connect to the unstable manifold from the other side in a point far away from the saddle. This structure is called $R_1R_2$ ring and is marked in blue colour in the plots. The major axis of the $R_1R_2$ still rotates in negative orientation with increasing bar length $a$. When the orientation of the major axis of the $R_1R_2$ ring approaches the $x$-axis, then the rings break and the unstable manifolds form open spirals as seen for $a \in [7.5,10.0]$. In our model we do not find persistent rings with the major axis pointing into the bar direction, so called $R_2$ rings. At the moment it is not yet clear to us what exactly are the properties of the model or the particular values of the involved parameters which are responsible for the absence of $R_2$ rings. Perhaps if we vary the values of other dynamical quantities (i.e., the mass or the angular velocity of the bar) $R_2$ rings might be possible but this is out of the scope of the present paper.

\begin{figure*}
\centering
\resizebox{\hsize}{!}{\includegraphics{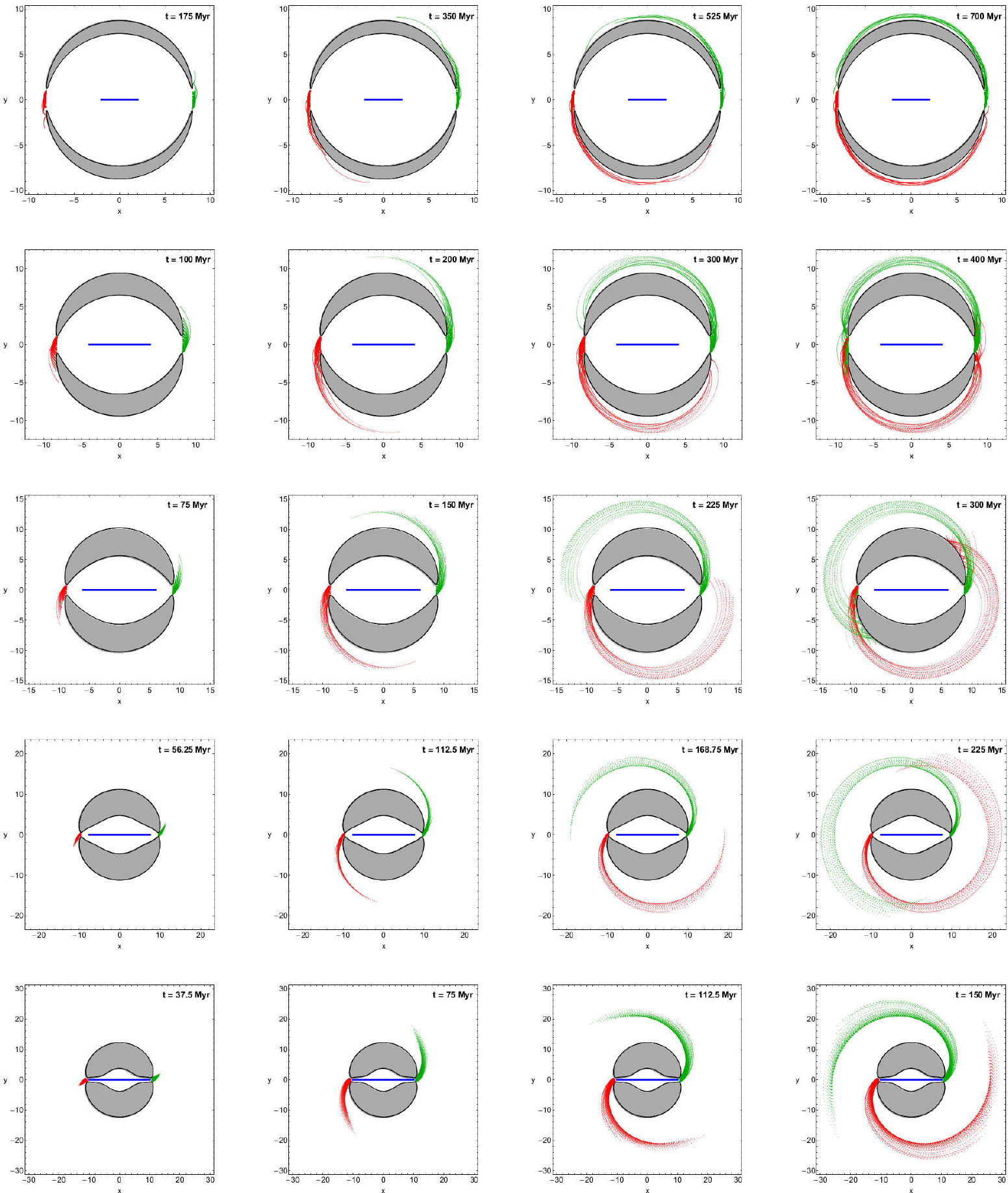}}
\caption{The distribution of the position of $10^6$ stars in the configuration $(x,y)$-plane initiated $(t = 0)$ within the Lagrange radius, for $\widehat{C} = 0.001$ and $\Omega_{\rm b} = -4.5$. The green segment contains stars that escaped through $L_2$, while the red segment contains stars that escaped through $L_3$. The horizontal blue lines in the interior region indicate the total length of the bar. (first row): $a = 2$; (second row): $a = 4$; (third row): $a = 6$; (forth row): $a = 7.5$; (fifth row): $a = 10$. (\textit{For the interpretation of references to color in this figure caption and the corresponding text, the reader is referred to the electronic version of the article.})}
\label{sr}
\end{figure*}

The classification of the morphologies shown in Fig. \ref{col} was done by eye inspection following the method according to which observers classify real galaxies. Generally it is rather easy to differentiate between the different types of the morphologies. However there are some borderline cases, such as between $R_1$ and $R_1'$ or between $R_1'$ and $R_1R_2$ for which the classification is a matter of personal judgment. Looking at the classification of the morphologies presented in the collection of Fig. \ref{col} we conclude that the shape of the stellar structures is directly related with the bar's semi-major axis. In particular, when the galactic bar is too small (or weak) the outer branches of the unstable manifolds form $R_1$ rings and $R_1'$ pseudo-rings, while on the other hand when we have the scenario of an elongated strong bar we observe the presence of $R_1R_2$ rings and open spirals. For additional useful information on the ring morphology see \citet{BC96}. Our results regarding the morphologies of the manifolds are completely consistent with that reported in \citet{ARGM09} and \citet{ARGBM09} where the galactic bar was modeled by a variety of potentials such as the Ferrers' \citep{F77}, or ad hoc potentials like the Dehnen \citep{D00} and the Barbanis-Woltjer \citep{BW67}.

\section{The fate of escaping stars}
\label{fate}

\begin{figure*}
\centering
\resizebox{\hsize}{!}{\includegraphics{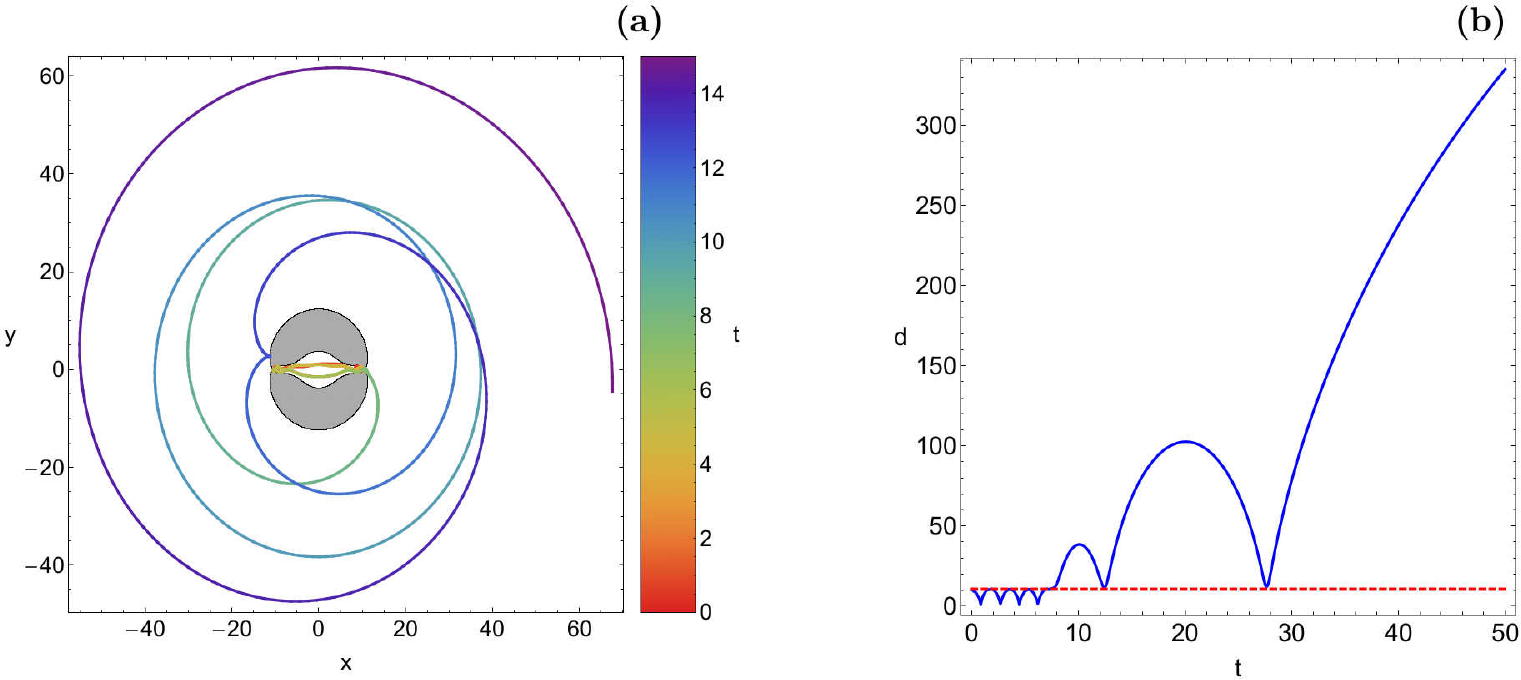}}
\caption{(a-left): The path of an escaping orbit in the configuration $(x,y)$ space for the first 15 time units, where colour is used to indicate its parametrization by time. (b-right): Time-evolution of the distance $d = \sqrt{x^2 + y^2}$ from the origin. The horizontal, red, dashed line indicates the Lagrange radius. (\textit{For the interpretation of references to color in this figure caption and the corresponding text, the reader is referred to the electronic version of the article.})}
\label{spr}
\end{figure*}

\begin{figure*}
\centering
\resizebox{\hsize}{!}{\includegraphics{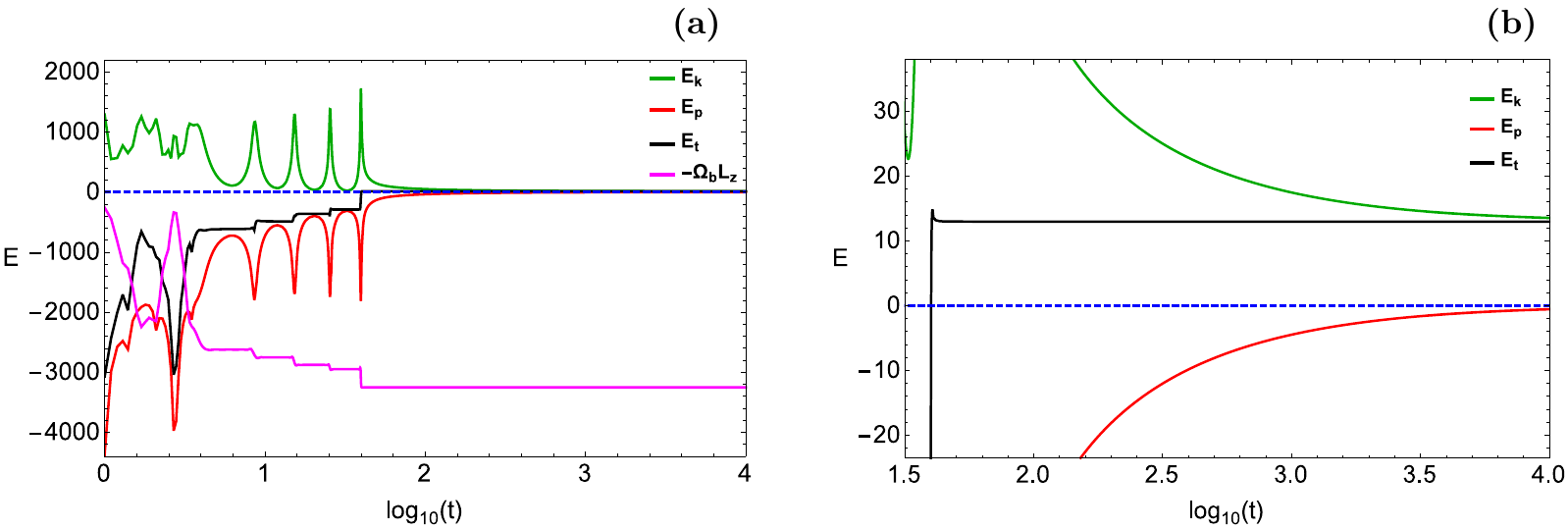}}
\caption{Time development of the kinetic energy $E_k$ (green), the potential energy $E_P$ (red), the total energy $E_t$ (black) and the quantity $- \Omega_b L_z$ (magenta) of the particle measured by an observer in the inertial frame. (b-right): A magnification of panel (a). (\textit{For the interpretation of references to color in this figure caption and the corresponding text, the reader is referred to the electronic version of the article.})}
\label{ener}
\end{figure*}

In Hamiltonian systems with open ZVCs test particles are free to escape from the system. In the potential interior there is a chaotic invariant set which directs the general flow. In this sense the general dynamics contains transient chaos where general orbits follow the chaotic motion of the invariant set for a finite time. Generally escaping orbits form complicated structures whose particular shapes are related to the dynamical system. For example stars which escape from tidally limited star clusters form stellar structures known as ``tidal tails" or ``tidal arms" \citep[see e.g.,][]{CMM05,dMCM05,JBPE09,KMH08,KKBH10}. Similarly escaping stars in barred galaxies have the tendency to create rings or spiral arms \citep[see e.g.,][]{EP14,GKC12,MT97,MQ06,QDB11,RDQ08,SK91} due to non-axisymmetric perturbations. The two spiral arms are developed from the two ends of the galactic bar and during galactic time-evolution they wind up around the banana-shaped forbidden regions of motion which surround the Lagrange points $L_4$ and $L_5$.

In Paper I we proved that our new gravitational galactic model has the ability to realistically model the formation as well as the time-evolution of the twin spiral structures observed in barred galaxies. Now we would like to examine if it can also model the creation of ring-shaped structures. As in all previous sections the variable parameter will be the bar's semi-major axis $(a \in [0.5, 10])$, while the values of all the other parameters remain constant according to SM. Usually in galactic simulations the bar rotates counter-clockwise, i.e., in direct sense with respect to the rotation of the galaxy itself therefore, the angular velocity of the bar should be $\Omega_b = -4.5$.

We modified our numerical integration code in order to record the output of all orbits in a three-dimensional grid of size $N_x \times N_y \times N_{p_x} = 100 \times 100 \times 100$, when $z_0 = p_{z_0} = 0$, while both signs of $p_{y_0}$ are allowed. This uniformly dense grid of initial conditions is centered at the origin of the configuration $(x,y)$ space, within the Lagrange radius $r_L = x(L_2)$. Our aim is to monitor the time-evolution of the escaping orbits and observe their final morphology.

In Fig. \ref{sr} we present the time-evolution of the position of the stars for five different values of the bar's semi-major axis $a$. The density of the points along one star orbit is taken to be proportional to the velocity of the star according to Paper II. Being more specific, a point is plotted (showing the position of a star on the configuration space), if an integer counter variable which is increased by one at every step of the numerical integration, exceeds the velocity of the star. Following this numerical technique we can simulate, in a way, a real $N$-body simulation of a barred galaxy, where the density of the stars will be highest where the corresponding velocity is lowest. It is seen that initially the vast majority of the stars are still inside the interior region. However, a small portion of stars escape through the Lagrange points $L_2$ (green) and $L_3$ (red) thus indicating the formation of spiral arms. As time goes by the two symmetrical arms grow in size and the final stellar structure starts to form. We observe that the time needed for the final structure to be developed decreases with increasing value of the bar's semi-major axis. The morphology of the final stellar structure depends on the specific value of $a$. In particular, for $a = 2$ a $R_1$ is formed, for $a = 4$ a $R_1'$ pseudo-ring is present, for $a = 6$ and 7.5 we have the scenario of a $R_1R_2$ ring, while for $a = 10$ twin open spiral arms are developed. It should be noted that the final stellar structure in all five cases coincide to that derived earlier in Section \ref{loman} for the unstable manifolds. Therefore we may conclude that the morphology of the final stellar structure (rings or open spirals) strongly depends on the specific value of the bar's semi-major axis but not on the distribution of the initial conditions of the orbits (orbits with initial conditions near the unstable Lyapunov orbits or uniformly spread across all over the interior region). Additional numerical simulations reveal that similar stellar structures are developed for other (lower or higher) values of total orbital energy.

Our last task will be to present an orbit which demonstrates very nicely the escape mechanism in our system when $\widehat{C} = 0.001$. It is one starting very close to the Lyapunov orbit $LO_2$, with initial conditions: $x_0 = 10.65, y_0 = p_{x0} = 0, p_{y0} > 0$. Fig. \ref{spr} shows the orbit in position space where part (a) presents the orbit in the $(x,y)$ plane and colour is used to indicate its parametrization by time, part (b) gives the distance $d = \sqrt{x^2 + y^2}$ from the origin as function of time. Part (a) of Fig. \ref{spr} is plotted in the coordinates of the rotating frame and we observe how the particle spirals around the inner region as soon as the distance becomes large. Fig. \ref{ener} shows the time development of the energy of the particle measured by an observer in the inertial frame, part (b) gives a magnification of part (a) showing in better detail the approach to the asymptotic behaviour for time to infinity. The kinetic energy of the particle $E_k$ is plotted in green, the potential energy $E_p$ in red and the total energy $E_t = E_k + E_p$ in black. The value 0 of the energy, i.e. the threshold for escape to infinity, is marked as blue, dashed curve because of its enormous importance for the asymptotic dynamics. It should be pointed out that the value of the Jacobi integral of motion (the total orbital energy in the rotating frame) is conserved during the numerical integration of the orbit shown in Fig. \ref{spr}. The value of the Jacobi quantity, i.e. the value $E_{\rm rot}$ of the Hamiltonian in the rotating frame, is conserved along any orbit. Then the combination of equations (\ref{Veff}) and (\ref{ham}) shows that the particle energy in the inertial frame $E_t$ can only change because of a corresponding change of the angular momentum $L_z$, we have $E_t = E_{\rm rot} + \Omega_b L_z$. Therefore it is instructive to include into Fig. \ref{ener} also the curve of the quantity $- \Omega_b L_z$ along the orbit,
i.e. as a function of time. We clearly see how the combination $E_t - \Omega_b L_z$ always stays constant at the initial
value $E_{\rm rot} = -3239.52940276$ ($\widehat{C} = 0.001$).

The orbit starts in the neighbourhood of the Lyapunov orbit $LO_2$ and leaves this neighbourhood close to the inner branch of the unstable manifold $W^u(LO_2)$. It moves along the bar to the neighbourhood of the opposite Lagrange point $L_3$ turns around and moves back to the vicinity of $L_2$, then makes another complete loop in the interior. Up to this point the orbit behaves very similar to a x1 orbit and it is also close to a heteroclinic orbit between the Lyapunov orbits $LO_2$ and $LO_3$ moving through the interior part of the potential well. After having made two complete oscillations along the bar, the orbit crosses the saddle at $L_2$ and this time leaves the saddle region along the outer branch of $W^u(LO_2)$. From here on it moves outside of the Lagrange radius and never manages to enter again the interior potential well. Orbits with different initial conditions are able to find their way back to the potential interior. In the outer part of the potential the orbit moves along spirals where for a while the distance from the origin oscillates. However, every time when the orbit comes close to the Lagrange radius it comes under the influence of the moving bar and gets a kick by the moving bar potential. In our particular case the phase of the bar potential is always such that the particle is accelerated by the bar potential and its total energy is increased. For other initial conditions the phases are different and the particle might get a kick pointing to the other side and decreasing the energy of the particle. Thereby the particle might even come back to the interior potential well. In a description in the rotating frame the kick applied by the bar is contained in the Coriolis forces in a more indirect form.

For our particular orbit under study we see very well in the black curve in Fig. \ref{ener}(a) how starting from $t \approx 8$ the energy increases in steps where the moment of the steps coincides with the closest approach to the origin and therefore with the closest approach to the bar, i.e. with the moment of the kick applied by the bar. In the magnification in Fig. \ref{ener}(b) we see that the last kick at a time $\log_{10}(t) \approx 1.6$ brings the total energy to a positive value. From here on the particle moves in the inertial frame out along a hyperbolic orbit determined by the long range part of the gravitational potential given as $ \Phi_{as} = - G M_{\rm total} / R$ where $M_{\rm total}$ is the total mass of the galaxy. The deviation of the bar potential from rotational symmetry is a quadrupole effect which in the potential goes to zero like $R^{-3}$. Therefore the energy of the particle remains constant as soon as it no longer comes close to the Lagrange radius and no longer feels the kicks by the bar. In the rotating frame the hyperbolic orbit of the inertial frame appears as spiral.

This orbit also demonstrates very well the basic ideas of chaotic scattering as an example of transient chaos. There is a chaotic invariant set containing a countable set of unstable periodic orbits, homoclinic and heteroclinic orbits to these periodic orbits and an over-countable set of truly chaotic orbits. The typical orbits do not belong to this particular subset of orbits. However, typical orbits with initial conditions close to stable manifolds of the chaotic invariant set move close to elements of the chaotic set for a finite time and thereby show for a finite time complicated behaviour which appears very similar to the true chaos exercised by the chaotic invariant set. The typical orbits finally find their way out of the neighbourhood of the chaotic set and switch to the simple asymptotic behaviour. Of course, because of time reversal invariance of Hamiltonian systems the same considerations also apply to the behaviour in the past direction of the dynamics even though this time direction is less important for applications to galaxies where usually the stars do not come from the outside but are created within the galaxy. What we have just described briefly is the typical mechanism of chaotic scattering, for more details see \citet{LT11}.

\section{Discussion and conclusions}
\label{disc}

The topic of the present article is a analytical gravitational multi-component model for barred galaxies which is simpler than the traditional models and which therefore allows extensive numerical studies with moderate computational resources.

An important part of the work in this article is the investigation how the dynamics of the model depends on the semi-major axis $a$ of the bar. We have studied the restriction of the dynamics to the invariant plane $z = p_z = 0$ which forms a 2 degree of freedom system. For energies below the escape threshold we found that the x1 orbits oscillate along the bar which form the bar and keep it stable. The important conclusions are that for $a$ approaching the value 10 the only large scale stable motion is the one with large negative $L$ and axis orientation perpendicular to the bar. The x1 orbits along the bar become unstable and either they become wide in the direction perpendicular to the bar or they disappear when the bar length $a$ approaches 10 and the energy approaches the escape threshold $E(L_2)$.

The escape dynamics of the dynamical barred galaxy model was revealed by integrating sets of initial conditions in several types of planes. In particular, for the configuration $(x,y)$ and the phase $(x,p_x)$ space we chose some characteristic energy levels above the energy of escape. In the $(x,\widehat{C})$ and $(y,\widehat{C})$ planes on the other hand, we used a continuous spectrum of energy values. Several well-defined basins of escape were found to coexist with highly fractal boundaries, while regions of bounded regular motion are also present in all type of planes. The escape dynamics of our dynamical model is very different with respect to that reported in Paper II, where using a simpler model it was found that the vast majority of all planes was covered by either non-escaping regular orbits or trapped chaotic orbits. Furthermore, the basins of escape were correlated with the corresponding escape times of the orbits. Our results indicate that the overall escape dynamics of our new model is very interesting and by all means realistic. For the numerical integration of the initial conditions of the orbits in each type of plane, we needed between about 0.5 hour and 4 days of CPU time on a Quad-Core i7 2.4 GHz PC, depending of course on the escape times of orbits in each case.

The Lyapunov orbits $LO_2$ and $LO_3$ are the central element of the chaotic invariant set and their stable and unstable manifolds direct the flow between the interior and the exterior of the central potential well. In the 3 dimensional phase space of the flow for fixed energy $W^s(LO_2)$ and $W^u(LO_2)$ form 2 dimensional tubes. The interior of the inner branches of $W^s(LO_2)$ contains the orbits heading towards the exits and correspondingly $W^s(LO_2)$ forms the boundaries of the basins of escape. Accordingly their stable manifolds delimit the basins of escape and their unstable manifolds determine the structure of the regions heavily influenced by the outgoing orbits. These outgoing orbits cause the perturbations forming the rings and spirals in the outer regions of the barred galaxy. The dependence of the resulting pattern of rings or spirals on the length $a$ of the bar coincides with the results of more complicated models.

Our last numerical task was to explore what happens to stars escaping from the barred galaxy. Do they move randomly or their orbits follow specific paths outside the central region of the galaxy? In order to answer this question we defined uniformly sets of initial conditions of orbits inside the central region and we monitored their time evolution. It was observed that escaping orbits through the saddle Lagrange points create interesting stellar structures (i.e., rings or open spirals). From the previous explanations it is clear that the morphology of these stellar structures strongly depends on the bar's semi-major axis and also that it does not depend on the distribution of initial conditions of the orbits. Numerical calculations lead to the same structures when either the orbits are started close to the Lyapunov orbit or when they are spread uniformly all over the interior region. It was proved that weak bars form $R_1$ rings $(0.5 \leq a \leq 3.0)$ and $R_1'$ pseudo-rings $(3.0 < a \leq 5.0)$, while strong elongated bars on the other hand favour the formation of $R_1R_2$ rings $(5.0 < a \leq 7.5)$ and open spirals $(a > 7.5)$.

Therefore the conclusion is that our dynamical model presents all the features observed in barred galaxies and expected for realistic models. We hope that the outcomes of our numerical investigation are useful in the field of escape dynamics in barred galaxies. Our results are considered as as a promising step in the task of understanding the properties of the escape mechanism of stars in galaxies with barred structure. In the second paper of the series we shall numerically explore the orbital as well as the escape dynamics of the full three-dimensional (3D) system.

\section*{Acknowledgments}

One of the authors (CJ) thanks DGAPA for financial support under grant number IG-101113. We would like to express our warmest thanks to the anonymous referee for the careful reading of the manuscript and for all the apt suggestions and comments which allowed us to improve both the quality and the clarity of our paper.

\bsp
\label{lastpage}


\begin{thebibliography}{}

\bibitem[\protect\citeauthoryear{Abraham \& Shaw}{1992}]{AS92} Abraham R.H., Shaw C.D., 1992, Dynamics, The Geometry of Behavior 2nd ed, Addison Wesley Redwood City

\bibitem[\protect\citeauthoryear{Aguirre et al.}{2001}]{AVS01} Aguirre J., Vallego J.C., Sanju\'{a}n M.A.F., 2001, Phys. Rev. E, 64, 066208

\bibitem[\protect\citeauthoryear{Aguirre \& Sanju\'{a}n}{2003}]{AS03} Aguirre J., Sanju\'{a}n M.A.F., 2003, Phys. Rev. E, 67, 056201

\bibitem[\protect\citeauthoryear{Aguirre et al.}{2009}]{AVS09} Aguirre J., Viana R.L., Sanju\'{a}n M.A.F., 2009, Rev. Mod. Phys., 81, 333

\bibitem[\protect\citeauthoryear{Athanassoula et al.}{1983}]{ABMP83} Athanassoula E., Bienaym\'{e} O., Martinet L., Pfenniger D., 1983, A\&A, 127, 349

\bibitem[\protect\citeauthoryear{Athanassoula et al.}{2009a}]{ARGM09} Athanassoula E., Romero-G\'{o}mez M., Masdemont J.J., 2009a, MNRAS, 394, 67

\bibitem[\protect\citeauthoryear{Athanassoula et al.}{2009b}]{ARGBM09} Athanassoula E., Romero-G\'{o}mez M., Bosma, A., Masdemont J.J., 2009b, MNRAS, 400, 1706

\bibitem[\protect\citeauthoryear{Athanassoula et al.}{2010}]{ARGBM10} Athanassoula E., Romero-G\'{o}mez M., Bosma A., Masdemont J.J., 2010, MNRAS, 407, 1433

\bibitem[\protect\citeauthoryear{Athanassoula et al.}{2011}]{ARGM11} Athanassoula E., Romero-G\'{o}mez M., Masdemont J.J., 2011, Memorie della Societa Astronomica Italiana Supplementi, 18, 97

\bibitem[\protect\citeauthoryear{Barbanis \& Woltjer}{1967}]{BW67} Barbanis B., Woltjer L., 1967, ApJ, 150, 461

\bibitem[\protect\citeauthoryear{Barrio et al.}{2008}]{BBS08} Barrio R., Blesa F., Serrano S., 2008, Europhys. Lett., 82, 10003

\bibitem[\protect\citeauthoryear{Benet et al.}{1996}]{BTS96} Benet L., Trautman D., Seligman T., 1996, Celest. Mech. Dyn. Astron., 66, 203

\bibitem[\protect\citeauthoryear{Benet et al.}{1998}]{BST98} Benet L., Seligman T., Trautman D., 1998, Celest. Mech. Dyn. Astron., 71, 167

\bibitem[\protect\citeauthoryear{Binney \& Tremaine}{2008}]{BT08} Binney J., Tremaine S., 2008, Galactic Dynamics, Princeton Univ. Press, Princeton, USA

\bibitem[\protect\citeauthoryear{Bleher et al.}{1988}]{BGOB88} Bleher S., Grebogi C., Ott E., Brown R., 1998, Phys. Rev. A, 38, 930

\bibitem[\protect\citeauthoryear{Bleher et al.}{1989}]{BOG89} Bleher S., Ott E., Grebogi C., 1989, Phys. Rev. Let., 63, 919

\bibitem[\protect\citeauthoryear{Bleher et al.}{1990}]{BGO90} Bleher S., Grebogi C., Ott E., 1990, Physica D, 46, 87

\bibitem[\protect\citeauthoryear{Blesa et al.}{2012}]{BSBS12} Blesa F., Seoane J.M., Barrio R., Sanju\'{a}n M.A.F., 2012, Int. J. Bifurc. Chaos, 22, 1230010

\bibitem[\protect\citeauthoryear{Bournaud \& Combes}{2001}]{BC02} Bournaud F., Combes F., 2002, A\&A, 392 83

\bibitem[\protect\citeauthoryear{Buta \& Combes}{1996}]{BC96} Buta R., Combes F., 1996, Fundamentals of Cosmic Physics, 17 95

\bibitem[\protect\citeauthoryear{Capuzzo Dolcetta et al.}{2005}]{CMM05} Capuzzo Dolcetta R., Di Matteo P., Miocchi, P., 2005, AJ, 129, 1906

\bibitem[\protect\citeauthoryear{Caranicolas \& Zotos}{2010}]{CZ10} Caranicolas N.D., Zotos E.E., 2010, Astronomische Nachrichten, 331, 330

\bibitem[\protect\citeauthoryear{Caranicolas \& Zotos}{2011}]{CZ11} Caranicolas N.D., Zotos E.E., 2011, Research in Astron. Astrophys., 11, 811

\bibitem[\protect\citeauthoryear{Chirikov}{1979}]{C79} Chirikov B.V., 1979, Phys. Rep., 52, 263

\bibitem[\protect\citeauthoryear{Contopoulos \& Papayannopoulos}{1980}]{CP80} Contopoulos G., Papayannopoulos T., 1980, A\&A, 92, 33

\bibitem[\protect\citeauthoryear{Darriba et al.}{2012}]{DMCG12} Darriba L.A., Maffione N.P., Cincotta P.M., Giordano C.M., 2012, International Journal of Bifurcation and Chaos, 22, 1230033

\bibitem[\protect\citeauthoryear{Dehnen}{2000}]{D00} Dehnen W., 2000, AJ, 119, 800

\bibitem[\protect\citeauthoryear{de Moura \& Letelier}{2000}]{dML00} de Moura A.P.S., Letelier P.S., 2000, Phys. Rev. E, 62, 4784

\bibitem[\protect\citeauthoryear{de Vaucouleurs}{1963}]{dV63} de Vaucouleurs G., 1963, ApJSS, 8, 31

\bibitem[\protect\citeauthoryear{Di Matteo et al.}{2005}]{dMCM05} Di Matteo P., Capuzzo Dolcetta R., Miocchi P., 2005, CeMDA, 91, 59

\bibitem[\protect\citeauthoryear{Ernst \& Peters}{2014}]{EP14} Ernst A., Peters T., 2014, MNRAS, 443, 2579 (Paper II)

\bibitem[\protect\citeauthoryear{Eskridge et al.}{2000}]{Ee00} Eskridge P.B., Frogel J.A., Pogge R.W., Quillen A.C., Davies R.L., DePoy D.L., Houdashelt M.L., et al., 2000, AJ, 119, 356

\bibitem[\protect\citeauthoryear{Ferrers}{1877}]{F77} Ferrers N.M., 1877, Q. J. Pure Appl. Math., 14, 1

\bibitem[\protect\citeauthoryear{G\'{o}mez et al.}{2000}]{GKL04} G\'{o}mez G., Koon W.S., Lo M.W., Marsden J.E., Masdemont J.J., Ross S.D., 2004, Nonlinearity, 17, 1571

\bibitem[\protect\citeauthoryear{Grand et al.}{2012}]{GKC12} Grand R.J.J., Kawata D., Cropper M., 2012, MNRAS, 421, 1529

\bibitem[\protect\citeauthoryear{H\'{e}non}{1969}]{H69} H\'{e}non M., 1969, A\&A, 1, 223

\bibitem[\protect\citeauthoryear{Innanen}{1980}]{I80} Innanen K.A., 1980, AJ, 85, 81

\bibitem[\protect\citeauthoryear{Ioka et al.}{2000}]{ITN00} Ioka K., Tanaka T., Nakamura T., 2000, ApJ, 528, 51

\bibitem[\protect\citeauthoryear{Jung \& Scholz}{1987}]{JS87} Jung C., Scholz H.J., 1987, J. Phys. A, 20, 3607

\bibitem[\protect\citeauthoryear{Jung \& Pott}{1989}]{JP89} Jung C., Pott S., 1989, J. Phys. A, 22, 2925

\bibitem[\protect\citeauthoryear{Jung et al.}{1995}]{JMS95} Jung C., Mejia-Monasterio C., Seligman T.H., 1995, Phys. Lett. A, 198, 306

\bibitem[\protect\citeauthoryear{Jung et al.}{1999}]{JLS99} Jung C., Lipp C., Seligman T.H., 1999, Ann. Phys., 275, 151

\bibitem[\protect\citeauthoryear{Jung \& Zotos}{2015}]{JZ15} Jung Ch., Zotos, E.E., 2015, PASA, 32, e042 (Paper I)

\bibitem[\protect\citeauthoryear{Just et al.}{2009}]{JBPE09} Just A., Berczik P., Petrov M., Ernst A., 2009, MNRAS, 392, 969

\bibitem[\protect\citeauthoryear{Koon et al.}{2000}]{KLMR00} Koon W.S., Lo M.W., Marsden J.E., Ross S.D., 2000, Chaos, 10, 427

\bibitem[\protect\citeauthoryear{K\"{u}pper et al.}{2008}]{KMH08} K\"{u}pper A.H.W., Macleod A., Heggie D.C., 2008, MNRAS, 387, 1248

\bibitem[\protect\citeauthoryear{K\"{u}pper et al.}{2010}]{KKBH10} K\"{u}pper A.H.W., Kroupa P., Baumgardt H., Heggie D.C., 2010, MNRAS, 401, 105

\bibitem[\protect\citeauthoryear{Lai et al.}{1993}]{LGB93} Lai Y.-C., Grebogi C., Bl\"{u}mel R., Kan I., 1993, Phys. Rev. Let., 71, 2212

\bibitem[\protect\citeauthoryear{Lai et al.}{2000}]{LMG00} Lai Y.-C., de Moura A.P.S., Grebogi C., 2000, Phys. Rev. E, 62, 6421

\bibitem[\protect\citeauthoryear{Lai \& T\'{e}l}{2011}]{LT11} Lai Y.-C., T\'{e}l T., 2011, Transient chaos, Springer New York

\bibitem[\protect\citeauthoryear{Lau et al.}{1991}]{LFO91} Lau Y.-T., Finn J.M., Ott E., 1991, Phys. Rev. Let., 66, 978

\bibitem[\protect\citeauthoryear{Lyapunov}{1949}]{L49} Lyapunov A., 1949, Ann. Math. Stud., 17

\bibitem[\protect\citeauthoryear{Masset \& Tagger}{1997}]{MT97} Masset F., Tagger M., 1997, A\&A, 322, 442

\bibitem[\protect\citeauthoryear{Masters et al.}{2011}]{MNH11} Masters K.L., Nichol R.C., Hoyle B., Lintott C., Bamford S.P., Edmondson E.M., Fortson L., et al. 2011, MNRAS, 411, 2026

\bibitem[\protect\citeauthoryear{McLin et al.}{2002}]{MSW02} McLin K.M., Stocke J.T., Weymann R.J., Penton S.V., Shull J.M. 2002, ApJ, 574, L115

\bibitem[\protect\citeauthoryear{Melvin \& Masters}{2013}]{MM13} Melvin T., Masters K., 2013, The Galaxy Zoo Team, Memorie della Societa Astronomica Italiana Supplementi, 25, 82

\bibitem[\protect\citeauthoryear{Minchev \& Quillen}{2006}]{MQ06} Minchev I., Quillen A.C., 2006, MNRAS, 368, 623

\bibitem[\protect\citeauthoryear{Miyamoto \& Nagai}{1975}]{MN75} Miyamoto W., Nagai R., 1975, PASJ, 27, 533

\bibitem[\protect\citeauthoryear{Olling \& Merrifield}{2000}]{OM00} Olling R.P., Merrifield M.R., 2000, MNRAS, 311, 361

\bibitem[\protect\citeauthoryear{Oppenheimer et al.}{2001}]{OHD01} Oppenheimer B.R., Hambly N.C., Digby A.P., Hodgkin S.T., Saumon D., 2001, Science, 292, 698

\bibitem[\protect\citeauthoryear{Penton et al.}{2002}]{PSS02} Penton S.V., Stocke J.T., Shull J.M., 2002, ApJ, 565, 720

\bibitem[\protect\citeauthoryear{Pfenniger}{1984}]{P84} Pfenniger D., 1984, A\&A 134, 373

\bibitem[\protect\citeauthoryear{Press}{1992}]{PTVF92} Press H.P., Teukolsky S.A, Vetterling W.T., Flannery B.P., 1992, Numerical Recipes in FORTRAN 77, 2nd Ed., Cambridge Univ. Press, Cambridge, USA

\bibitem[\protect\citeauthoryear{Quillen et al.}{2011}]{QDB11} Quillen A.C., Dougherty J., Bagley M.B., Minchev I., Comparetta J., 2011, MNRAS, 417, 762

\bibitem[\protect\citeauthoryear{Romero-G\'{o}mez et al.}{2006}]{RGMA06} Romero-G\'{o}mez M., Masdemont J.J., Athanassoula E., Garc\'{i}a-G\'{o}mez C., 2006, A\&A, 453, 39

\bibitem[\protect\citeauthoryear{Romero-G\'{o}mez et al.}{2007}]{RGAM07} Romero-G\'{o}mez M., Athanassoula E., Masdemont J.J., Garc\'{i}a-G\'{o}mez C., 2007, A\&A, 472, 63

\bibitem[\protect\citeauthoryear{Ro\v{s}kar et al.}{2008}]{RDQ08} Ro\v{s}kar R., Debattista V.P., Quinn T.R., Stinson G.S., Wadsley J., 2008, ApJ, 684, L79

\bibitem[\protect\citeauthoryear{Sellwood \& Kahn}{1991}]{SK91} Sellwood J.A., Kahn F.D., 1991, MNRAS, 250, 278

\bibitem[\protect\citeauthoryear{Sheth et al.}{2003}]{SRSS03} Sheth K., Regan M.W., Scoville N.Z., Strubbe L.E., 2003, ApJ, 592, L13

\bibitem[\protect\citeauthoryear{Sheth et al.}{2008}]{SEE08} Sheth K., Elmegreen D.M., Elmegreen B.G., Capak P., Abraham R.G., Athanassoula E., Ellis R.S., et al. 2008, ApJ, 675, 1141

\bibitem[\protect\citeauthoryear{Skokos}{2001}]{S01} Skokos C., 2001, Journal of Physics A, 34, 10029

\bibitem[\protect\citeauthoryear{Skokos et al.}{2002}]{SPA02} Skokos C., Patsis P. A., Athanassoula E., 2002a, MNRAS, 333, 847

\bibitem[\protect\citeauthoryear{Steidel et al.}{2002}]{SKS02} Steidel C.C., Kollmeier J.A., Shapley A.E., Churchill, C.W., Dickinson M., Pettini M., 2002, ApJ, 570, 526

\bibitem[\protect\citeauthoryear{Wechsler et al.}{2002}]{WBP02} Wechsler R.H., Bullock J.S., Primack J.R., Kravtsov A.V., Dekel A., 2002, ApJ, 568, 52

\bibitem[\protect\citeauthoryear{Wolfram}{2003}]{W03} Wolfram S., 2003, The Mathematica Book. Wolfram Media, Champaign

\bibitem[\protect\citeauthoryear{Zotos}{2012}]{Z12} Zotos E.E., 2012, Research in Astr. Astrophys., 12, 500

\bibitem[\protect\citeauthoryear{Zotos}{2014a}]{Z14a} Zotos E.E., 2014a, A\&A, 563, A19

\bibitem[\protect\citeauthoryear{Zotos}{2014b}]{Z14b} Zotos E.E., 2014b, Nonlinear Dynamics, 78, 1389

\bibitem[\protect\citeauthoryear{Zotos}{2015a}]{Z15a} Zotos E.E., 2015a, Nonlinear Dynamics, 82, 357

\bibitem[\protect\citeauthoryear{Zotos}{2015b}]{Z15b} Zotos E.E., 2015b, A\&SS, 358, 4

\bibitem[\protect\citeauthoryear{Zotos}{2015c}]{Z15c} Zotos E.E., 2015c, A\&SS, 367, 7

\bibitem[\protect\citeauthoryear{Zotos}{2015d}]{Z15d} Zotos E.E., 2015d, MNRAS, 446, 770

\bibitem[\protect\citeauthoryear{Zotos}{2015e}]{Z15e} Zotos E.E., 2015e, MNRAS, 452, 193

\end{thebibliography}
\end{document}